\documentclass[fleqn,12pt]{wlscirep}
\usepackage[utf8]{inputenc}
\usepackage[T1]{fontenc}
\usepackage{xcolor}
\usepackage[font=scriptsize]{caption}
\usepackage[version=4]{mhchem}
\usepackage{pifont}%http://ctan.org/pkg/pifont
\usepackage{chemformula}
\usepackage[left]{lineno}

\usepackage{chemfig}

\title{OH as a probe of the warm water cycle in planet-forming disks}

\author[1*]{Marion Zannese}\author[1]{Benoît Tabone}\author[1]{Emilie Habart}\author[2]{Javier R. Goicoechea}\author[2]{Alexandre Zanchet}\author[3,4]{Ewine F. van Dishoeck}\author[5]{Marc C. van Hemert}
\author[6]{John H.~Black}\author[3,7]{Alexander G. G. M. Tielens}\author[8]{A. Veselinova}\author[8]{P. G. Jambrina}\author[9]{M. Menendez}\author[9]{E. Verdasco}\author[9]{F. J. Aoiz}\author[8]{L. Gonzalez-Sanchez}\author[1]{Boris Trahin}\author[10]{Emmanuel Dartois}\author[11]{Olivier Bern\'e}\author[12,13,14]{Els Peeters}\author[15,16,17]{Jinhua He}\author[12,13]{Ameek Sidhu}\author[12,13]{Ryan Chown}\author[11]{Ilane Schroetter}\author[18]{Dries Van De Putte}\author[11]{Am\'elie Canin}\author[19]{Felipe Alarc\'on}\author[1]{Alain Abergel}\author[19]{Edwin A. Bergin}\author[20,21]{Jeronimo Bernard-Salas}\author[22]{Christiaan Boersma}\author[23]{Emeric Bron}\author[12,13,14]{Jan Cami}
\author[24]{Daniel Dicken}\author[1]{Meriem Elyajouri}\author[25]{Asunci\'on Fuente}\author[18,26]{Karl D. Gordon}\author[11]{Lina Issa}\author[11]{Christine Joblin}\author[1]{Olga Kannavou}\author[12]{Baria Khan}\author[27]{Ozan Lacinbala}\author[23]{David Languignon}\author[28,29]{Romane Le Gal}\author[22]{Alexandros Maragkoudakis}\author[1,23]{Raphael Meshaka}\author[30]{Yoko Okada}\author[31,32]{Takashi Onaka}\author[12]{Sofia Pasquini}\author[19]{Marc W. Pound}\author[18,33]{Massimo Robberto}\author[34,35]{Markus Röllig}\author[12,13]{Bethany Schefter}\author[1,6]{Thi\'ebaut Schirmer}\author[36]{S\'ilvia Vicente}\author[7]{Mark G. Wolfire}

\affil[1]{Universit\'e Paris-Saclay, CNRS, Institut d’Astrophysique Spatiale, 91405 Orsay, France}
\affil[*]{marion.zannese@universite-paris-saclay.fr}
\affil[2]{Instituto de F\'{\i}sica Fundamental  (CSIC),  Calle Serrano 121-123, 28006, Madrid, Spain}
\affil[3]{Leiden Observatory, Leiden University, 2300 RA Leiden, The Netherlands}
\affil[4]{Max-Planck Institut für Extraterrestrische Physik (MPE), Giessenbachstr. 1, 85748 Garching, Germany}
\affil[5]{Leiden Institute of Chemistry, Gorlaeus Laboratories, Leiden University, Einsteinweg 55, 2333 CC Leiden, The Netherlands}
\affil[6]{Department of Space, Earth and Environment, Chalmers University of Technology, Onsala Space Observatory, 43992, Onsala,
Sweden}
\affil[7]{Astronomy Department, University of Maryland, College Park, MD 20742, USA}
\affil[8]{Departamento de Química Física, University of Salamanca, Plaza Caidos S/N, E-37008, Salamanca, Spain}
\affil[9]{Departamento de Química Física (Unidad Asociada al CSIC), Universidad Complutense de Madrid, Ciudad Universitaria, S/N,
E-20840, Madrid, Spain}
\affil[10]{Institut des Sciences Mol\'eculaires d'Orsay, CNRS, Universit\'e Paris-Saclay, 91405 Orsay, France}
\affil[11]{Institut de Recherche en Astrophysique et Plan\'etologie, Universit\'e Toulouse III - Paul Sabatier, CNRS, CNES, 9 Av. du colonel Roche, 31028 Toulouse Cedex 04, France}
\affil[12]{Department of Physics \& Astronomy, The University of Western Ontario, London ON N6A 3K7, Canada}
\affil[13]{Institute for Earth and Space Exploration, The University of Western Ontario, London ON N6A 3K7, Canada}
\affil[14]{Carl Sagan Center, SETI Institute, 339 Bernardo Avenue, Suite 200, Mountain View, CA 94043, USA}
\affil[15]{Yunnan Observatories, Chinese Academy of Sciences, 396 Yangfangwang, Guandu District, Kunming, 650216, China}
\affil[16]{Chinese Academy of Sciences South America Center for Astronomy, National Astronomical Observatories, CAS, Beijing 100101,  China}
\affil[17]{Departamento de Astronomica, Universidad de Chile, Casilla 36-D, Santiago, Chile}
\affil[18]{Space Telescope Science Institute, 3700 San Martin Drive, Baltimore, MD, 21218, USA }
\affil[19]{Department of Astronomy, University of Michigan, 1085 South University Avenue, Ann Arbor, MI 48109, USA}
\affil[20]{ACRI-ST, Centre d’Etudes et de Recherche de Grasse (CERGA), 10 Av. Nicolas Copernic, F-06130 Grasse, France}
\affil[21]{INCLASS Common Laboratory., 10 Av. Nicolas Copernic, 06130 Grasse, France}
\affil[22]{NASA Ames Research Center, MS 245-6, Moffett Field, CA 94035-1000, USA}
\affil[23]{LERMA, Observatoire de Paris, PSL Research University, CNRS, Sorbonne Universit\'es, F-92190 Meudon, France}
\affil[24]{UK Astronomy Technology Centre, Royal Observatory Edinburgh, Blackford Hill, Edinburgh EH9 3HJ, UK}
\affil[25]{ Centro de Astrobiolog\'{\i}a, CAB, CSIC-INTA, Carretera de Ajalvir Km 4, Torrej\'on de Ardoz, 28850, Madrid, Spain}
\affil[26]{Sterrenkundig Observatorium, Universiteit Gent, Gent, Belgium}
\affil[27]{Quantum Solid State Physics (QSP), Celestijnenlaan 200d - box 2414, 3001 Leuven, Belgium}
\affil[28]{Institut de Plan\'etologie et d'Astrophysique de Grenoble (IPAG), Universit\'e Grenoble Alpes, CNRS, F-38000 Grenoble, France}
\affil[29]{Institut de Radioastronomie Millim\'etrique (IRAM), 300 Rue de la Piscine, F-38406 Saint-Martin d'H\`{e}res, France }
\affil[30]{I. Physikalisches Institut der Universit\"{a}t zu K\"{o}ln, Z\"{u}lpicher Stra{\ss}e 77, 50937 K\"{o}ln, Germany}
\affil[31]{Department of Astronomy, Graduate School of Science, The University of Tokyo, 7-3-1 Bunkyo-ku, Tokyo 113-0033, Japan}
\affil[32]{Department of Physics, Faculty of Science and Engineering, Meisei University, 2-1-1 Hodokubo, Hino, Tokyo 191-8506, Japan}
\affil[33]{Johns Hopkins University, 3400 N. Charles Street, Baltimore, MD, 21218, USA}
\affil[34]{Physikalischer Verein – Gesellschaft für Bildung und Wissenschaft, Robert-Mayer-Str. 2, 60325 Frankfurt am Main, Germany}
\affil[35]{Institut für Angewandte Physik, Max-von-Laue-Str. 1, 60438 Frankfurt am Main, Germany}
\affil[36]{Instituto de Astrof\'isica e Ci\^{e}ncias do Espa\c co, Tapada da Ajuda, Edif\'icio Leste, 2\,$^{\circ}$ Piso, P-1349-018 Lisboa, Portugal}


\begin{abstract}
  \bf {Water is a key ingredient for the emergence of life as we know it. Yet, its destruction and reformation in space remains unprobed in warm gas. Here, we detect the hydroxyl radical (OH) emission from a planet-forming disk exposed to external far-ultraviolet (FUV) radiation with the \textit{James Webb Space Telescope}. The observations are confronted with the results of quantum dynamical calculations. The highly excited OH infrared rotational lines are the tell-tale signs of H$_2$O destruction by FUV. The OH infrared ro-vibrational lines are attributed to chemical excitation via the key reaction O+H$_2$$\rightarrow$OH+H which seeds the formation of water in the gas-phase. We infer that the equivalent of the Earth ocean's worth of water is destroyed per month and replenished. These results show that under warm and irradiated conditions water is destroyed and efficiently reformed via gas-phase reactions. This process, assisted by diffusive transport, could reduce the HDO/H$_2$O ratio in the warm regions of planet-forming disks. }
\end{abstract}

\begin{document}

\flushbottom
\maketitle

\thispagestyle{empty}

\bigskip
\bigskip
\noindent

Water is a key ingredient in the emergence of life and is, therefore, a key aspect in the assessment of the habitability of (exo)planets. Yet, the trail of water to planets remains unclear. In a disk's inner regions ($\lesssim 10~$au), where the terrestrial and sub-Neptune planets are expected to form, a significant fraction of the water inherited from cold clouds is destroyed and reformed\cite{vanDishoeck_2021,Glassgold_2009,Bethell_2009}.  In fact, theoretical models predict that under warm ($T>300~$K) and FUV ($6 < h \nu < 13.6~$eV) irradiated conditions present in disk’s atmospheres, water is destroyed and reformed by the chemical cycle\cite{vanDishoeck_2021,Glassgold_2009,Bethell_2009} $\text{O}  \rightleftarrows \text{OH} \rightleftarrows \text{H}_2\text{O}$. Notably, this cycle is one of the processes that could lower the deuterium-to-hydrogen enrichment of water inherited from cold clouds and explain the intermediate D/H ratio found in Earth's oceans\cite{Thi2010}.

In the past, understanding the detailed balance between the formation and destruction processes in a protoplanetary disk setting has been challenging because of the limited observational
information. Warm water has been detected in planet-forming disks with \textit{Spitzer}-IRS \cite{Salyk_2011,Carr_2011,Pontoppidan_2014} and ground-based instruments\cite {Mandell_2012} but its ongoing chemistry remains unprobed. Here, we unveil the water cycle under warm and irradiated conditions thanks to the combination of \textit{JWST} observations of OH and quantum dynamical calculations.

\begin{figure}
    \centering
    \includegraphics[width=\linewidth]{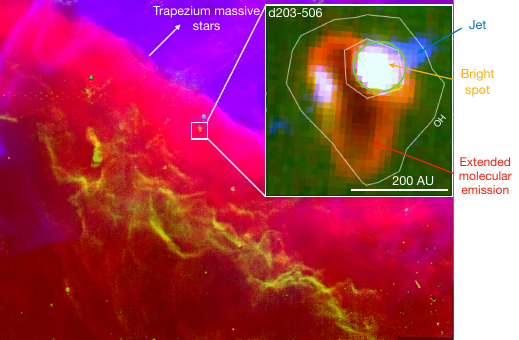}
    \caption{JWST NIRCam composite image of the Orion Bar, located in the Orion molecular cloud. Red is the 3.35$\mu$m emission (F335M NIRCam filter), blue is the emission of Pa$\alpha$ (F187N filter subtracted by F182M filter) and green is the emission of the H$_2$ 0-0 S(9) line at 4.70 $\mu$m (F470N filter subtracted by F480M filter).  The inset shows a zoom-in of the d203-506 planet-forming disk where OH lines are detected. Red is the emission of the H$_2$ 1-0 S(1) line at $2.12 ~\mu$m (F212N filter), blue is the [FeII] line emission at $1.64 ~\mu$m (F164N filter), and green is the emission in the F140M broad-band near-IR filter that traces scattered light around 1.4~$\mu$m (F140M broad band filter). The white contours represent the emission of the OH rotational line at 9.79 $\mu$m detected with MIRI-MRS (levels are 1.1, 2.5 $\times$10$^{-5}$ erg cm$^{-2}$ s$^{-1}$ sr$^{-1}$).
    The bright spot in the northwestern part of the d203-506 system coincides with the region of interaction between a jet and the photoevaporative wind. The brightest OH emission originates from here. The spectra shown in Fig. \ref{fig:MIRI_spectrum} and \ref{fig:NIRSpec_spectrum} are an average over the region delineated by the green circle in order to have the best S/N. Image credits: NASA, ESA, CSA, PDRs4All ERS Team, pdrs4all.org.}
    \label{fig:proplyd}
\end{figure}

The observational data stem from the \textit{JWST} Early Release Science (ERS) program ``PDRs4All'' \cite{ERS_2022}, which performed spectroscopic observations of the Orion Bar, an interstellar cloud exposed to the intense FUV radiation from the massive stars of the Trapezium cluster\cite{Tielens_1993}(FUV flux at the ionization front equal to $2-7 \times 10^4$ times that of the local interstellar medium\cite{Peeters_2023} (ISM)). The Solar System presumably formed in a similar environment in which external UV radiation influenced the temperature, mass budget, and chemical composition of the solar nebula\cite{Adams_2010}. Fig. \ref{fig:proplyd} shows the d203-506 disk, located in the observed area, as a dark lane against a bright background of the nebulae (see Methods for its detailed characteristics). 
A photoevaporative wind, seen as a more diffuse emission of H$_2$, is launched from the upper layers of the disk \cite{Winter_2022}. This is the result of the intense FUV radiation from the massive stars in the proximity of d203-506 that heats the gaseous disk.
In addition, the young star at the center launches a collimated high-velocity jet, seen in [FeII] line emission which carves out a cavity in the  wind. The North-West part of the cavity is particularly bright in excited molecular lines. This enhanced emission is likely due to a local increase in density induced by jet bow-shocks\cite{Tabone_2018}, and the direct irradiation from the Trapezium stars or the central star itself.
Here, we report the detection of infrared rotational and ro-vibrational emission of OH by both MIRI-MRS and NIRSpec, respectively. The OH emission is detected where there is bright H$_2$ emission, namely all over the photoevaporative wind, revealing active gas-phase oxygen chemistry fueled by warm H$_2$ (see also Extended Data Fig. \ref{fig:map_OH}). In this study, we focus on the bright spot to obtain high signal-to-noise spectra.

\begin{figure}
    \centering
    \includegraphics[width=\linewidth]{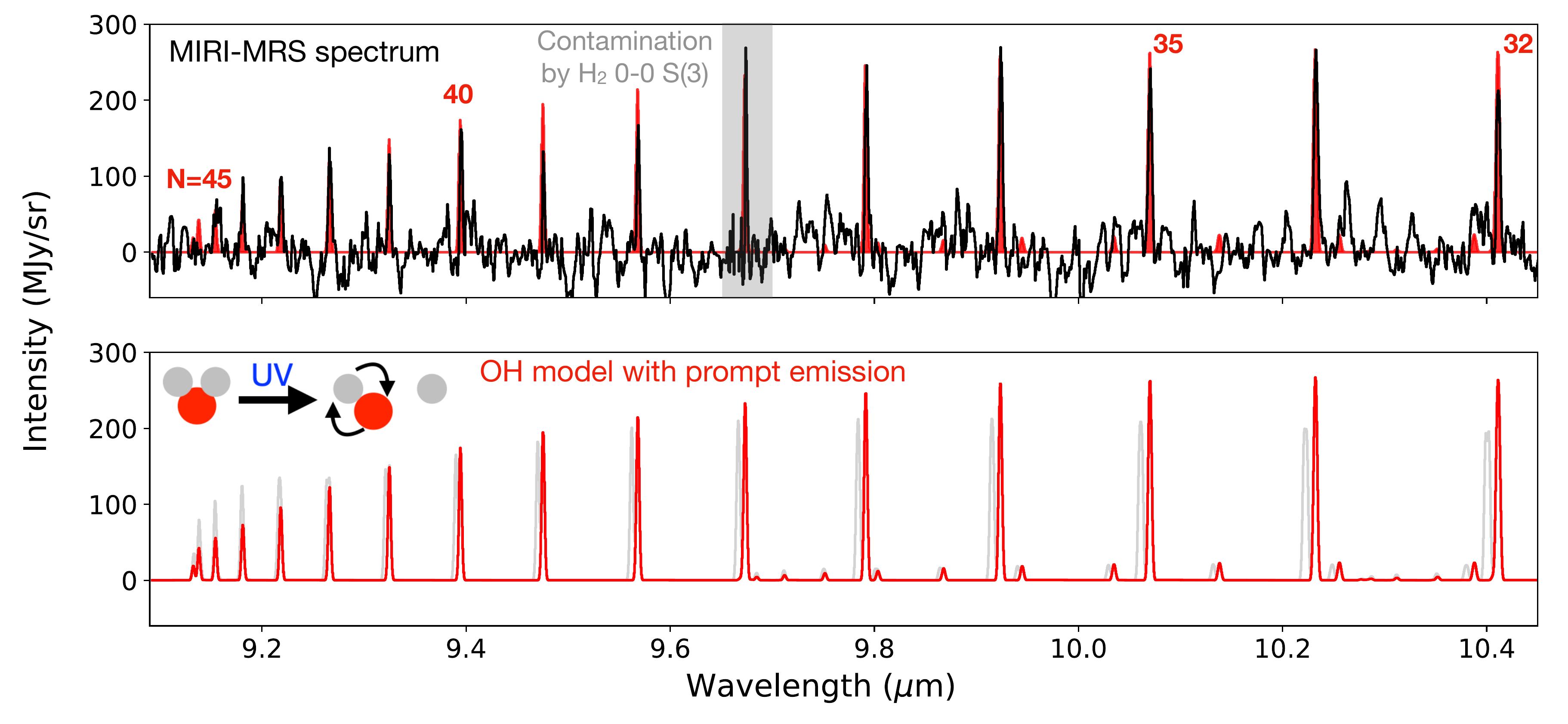}
    \caption{Evidence for H$_2$O photodissociation from OH rotational lines detected by MIRI-MRS. (Top panel) Observed MIRI-MRS spectrum where the dust continuum and bright lines, other than that from OH, have been subtracted (see Method). Rotationally excited OH lines are detected up to levels $N=44$. 
    (Bottom panel) Synthetic spectra from \texttt{GROSBETA}. The spectrum in red assumes that water photodissociation produces OH in symmetric $\Lambda$-doubling states, whereas the gray spectrum assumes equal distribution among the sub-levels.}
    \label{fig:MIRI_spectrum}
\end{figure}

\paragraph{Evidence for H$_2$O photodissociation.} The MIRI-MRS spectrum shown in Fig. \ref{fig:MIRI_spectrum} reveals a series of highly excited rotational OH lines in the ground vibrational state, corresponding to a change in rotational quantum number of $N$ $\rightarrow$ $N$-1. In total, lines from 23 rotational levels from $N=44$ down to $N=18$ are detected in the $9-13~\mu$m region, which 
probes upper-level energies as high as 45,000 K (see Extended Data Fig. \ref{fig:energy_levels}). The excitation diagram of OH reveals an extremely high excitation temperature of 10,000 K. This is ten times larger than the gas temperature inferred from the H$_2$ rotational lines (see Extended Data Fig. \ref{fig:diag_rot_h2} and \ref{fig:excitation_diagram}).
The detection of extremely rotationally excited OH is the smoking gun of water photodissociation.
As already acknowledged from previous \textit{Spitzer} observations\cite{Tappe_2008,Najita_2010,Carr_2014} based on seminal molecular physics studies \cite{Harich_2001,van_Harrevelt_2001}, the only process that can excite these lines is H$_2$O photodissociation, via its $\tilde{B}$ electronic state by short wavelength FUV photons forming OH in high-$N$ states. The levels that are directly populated by water photodissociation ($N \simeq 35-45$) were not detected before due to the limited spectral resolution and sensitivity of \textit{Spitzer}-IRS. Furthermore, the higher spectral resolution of MIRI-MRS allows us to separate the symmetric from the anti-symmetric component of each rotational line down to $9.4~\mu$m (see Fig. \ref{fig:MIRI_spectrum} and Extended Data Fig. \ref{fig:energy_levels}). Compared to the less excited lines observed with \textit{Spitzer} \cite{Carr_2014}, we find a much higher ratio ($\ge$10) between the symmetric and anti-symmetric lines. This is consistent with recent quantum calculations using a full-dimensional wave packet method which attributes this propensity to the $\tilde{B} \rightarrow X$ conical intersection pathway \cite{Zhou_2015}.

In order to analyze the MIRI-MRS spectrum, synthetic \textit{JWST} spectra were computed using the \texttt{GROSBETA} code\cite{Tabone_2021}, which takes into account the production of OH in excited states (see Method). As an input, we use the state distribution of nascent OH produced by H$_2$O photodissociation via its first two electronic states as computed by ref.\cite{van_Harrevelt_2000, van_Harrevelt_2001}. As shown in Fig. \ref{fig:MIRI_spectrum}, the model agrees well with the observed spectrum. In particular, we recover the steep increase in line intensity from $\lambda$ = 9.15 $\mu$m onward, and the relatively constant line intensities beyond $9.5~\mu$m. This behavior is intrinsically related to the rotational distribution of nascent OH (see Extended Data Fig. \ref{fig:state_distrib}). Once an OH molecule is formed in a rotationally excited state, it decays via the $N \rightarrow N-1$ radiative transitions, a process called ``radiative cascade''\cite{Tabone_2021}. 
The agreement between the model and the observations is not only an astronomical confirmation of a basic molecular process but, as shown in the next section, also provides access to the destruction rate of water and its local abundance.

\begin{figure}
    \centering
    \includegraphics[width=\linewidth]{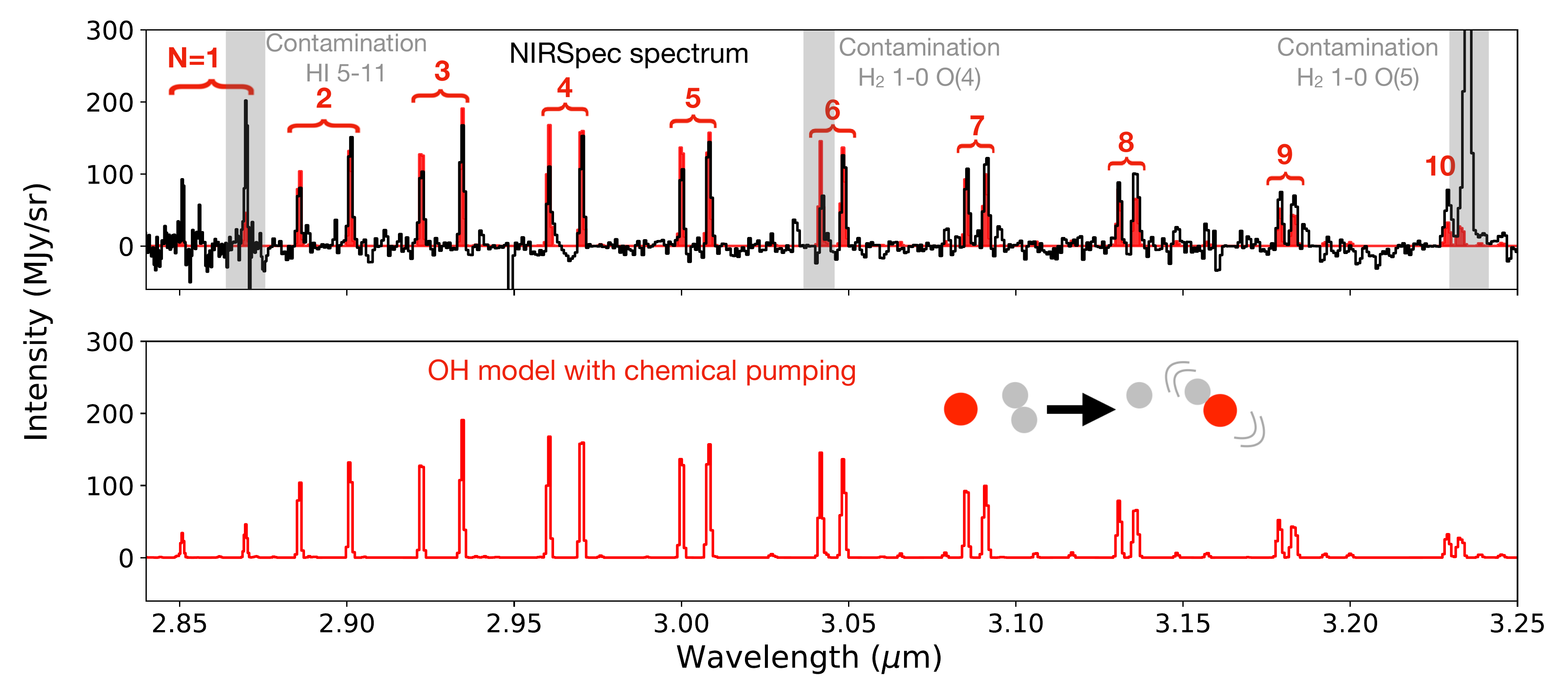}
    \caption{Evidence for formation-pumping via O+H$_2$ from OH ro-vibrational lines detected by NIRSpec. (Top panel) Observed NIRSpec spectrum where the dust continuum and the bright lines, other than those from OH, have been subtracted (see Method). Each ro-vibrational transition is split by the spin-orbit coupling, however the $\Lambda$-doubling is unresolved. 
    (Bottom panel) Synthetic spectra from \texttt{GROSBETA} including the excitation of OH via formation-pumping.}
    \label{fig:NIRSpec_spectrum}
\end{figure}

\paragraph{Evidence for chemical ("formation")-pumping by O+H$_2$.} The $v=1-0$ ro-vibrational lines of OH are detected at shorter near-infrared (near-IR) wavelengths by NIRSpec, with upper rotational quantum number from $N=1$ up to 10 (see Fig. \ref{fig:NIRSpec_spectrum} and Extended Data Fig. \ref{fig:energy_levels}). These transitions have been detected in very dense environments like the innermost regions of protoplanetary disks\cite{Mandell_2012} ($n_{\text{H}} \gtrsim 10^8~$cm$^{-3}$), but not in a lower density environment such as the outer layers of the d203-506 system. 
The excitation of the rotational levels within the $v =1$ state is well described by a single excitation temperature of about $1,000~$K (Extended Data Fig. \ref{fig:excitation_diagram}), close to the gas temperature inferred from H$_2$ emission (Extended Data Fig. \ref{fig:diag_rot_h2}) and ten times smaller than the excitation temperature of the mid-IR lines, implying a different origin.

We attribute the near-IR OH lines to formation-pumping via the reaction O+H$_2$, which is known to produce OH in vibrationally excited states \cite{Balakrishnan_2006}, but has hitherto never been observed in space. We amended the \texttt{GROSBETA} model to include the excitation of OH via the O+H$_2$ reaction using the state distribution of nascent OH extracted from the recent quantum calculations of ref.\cite{Veselinova_2021} (see Method and Extended Data Fig. \ref{fig:state_distrib}). Since the state distribution of nascent OH depends on both the gas temperature and the state distribution of H$_2(v,J)$, we used the observed lines of H$_2$ to compute, self-consistently, the distribution of nascent OH. The synthetic \texttt{GROSBETA} model shows good agreement with the NIRSpec spectrum (see Fig. \ref{fig:NIRSpec_spectrum}). 
The OH $v =2-1$ lines are also predicted to be an order of magnitude weaker, in line with the upper limit set by the observations; quantum calculations predict that $20\%$ of the OH is produced in the $v=1$ state, with less than $1\%$ in higher excited vibrational states ($v\ge 2$). The emission process is relatively simple, as an OH product formed in a $v=1$ state rapidly cascades down via the ro-vibrational transitions. Other processes are known to produce OH in vibrational states (UV pumping, H$_2$O photodissociation via its $\tilde{A}$ electronic state, IR pumping) but we find that only inelastic collisions could significantly contribute to the observed emission if OH is very abundant ($x($OH$) \gtrsim 10^{-5}$, see detailed discussion in Method). The good agreement between observations and model points to formation-pumping of OH, thereby supporting earlier proposal based on \textit{Spitzer}-IRS and \textit{Herschel} observations \cite{Goicoechea_2011,Carr_2014}.

\paragraph{Warm gas-phase oxygen chemistry in action} 
Because the newly formed OH products immediately decay via the observed transitions, the fluxes of these lines are directly proportional to the formation rate of OH via either route (see Table \ref{Table:parameters}, and Method for further details). 
From the mid-IR lines, we derive an amount of water photodissociated per unit of time of $\Phi =  1.1 \times 10^{10}$ cm$^{-2}$ s$^{-1}$. Considering the size of the bright spot 
(1.53$\times$10$^{-12}$ sr, at a distance of 414~pc \cite{Menten_2007}), this corresponds to, roughly, the equivalent of the Earth ocean's worth of water being photodestroyed per month\footnote{The oceans of the Earth are composed of about 5$\times$10$^{46}$ molecules of water.}. Similarly, from the near-IR lines, we infer an amount of OH formed via O+H$_2$ per unit of time of $R=1.2 \times 10^{11}$~cm$^{-2}$ s$^{-1}$.

These formation rates shed new light on the water cycle under warm and irradiated conditions. The mid-IR lines of OH reveal that water is actively photodestroyed in the d203-506 system. Water lines are not detected with MIRI-MRS, as predicted by non-LTE models (see Extended Data Fig. \ref{fig:H2O_GROSBETA}), likely due to the low density of the gas which is much lower than the critical densities of the H$_2$O transitions observable with \textit{JWST}. Yet, from the amount of H$_2$O photodissociated  per second, we infer a substantial amount of unseen H$_2$O in the warm gas with $N$(H$_2$O$) \simeq 1-5 \times 10^{15}$~cm$^{-2}$ (see Method). This quantity is still two orders of magnitude too low for water UV-shielding in this environment\cite{Bosman_2022}. As estimated from the local FUV radiation field, the destruction timescale of water is about a day, much shorter than the dynamical timescale of the outflowing gas, which is longer than half a century (see Method).
Therefore, the illuminated water cannot originate from inherited water formed in cold conditions. In fact, our analysis of the near-IR lines of OH demonstrates that the formation rate of OH via O+H$_2$ is an order of magnitude larger than the destruction rate of H$_2$O. Water is thus efficiently replenished by the gas-phase formation route initiated by the reaction O+H$_2 \rightarrow$ OH+H. Our estimates also indicate that only a small fraction of OH, about $\Phi/R \simeq 10 \%$, is effectively converted into H$_2$O. Notably, OH is also photodissociated, reducing the probability of OH being converted into H$_2$O. Interestingly, OH photodissociation has already been unveiled by the Hubble Space Telescope observations of [OI] emission at $\lambda = 6300~$\AA\cite{Bally_2000}, which is likely excited by OH photodissociation \cite{Storzer_1998}. Assuming that [OI] emission is produced via OH+$h \nu \rightarrow$ O+H with a 50$\%$ probability to form oxygen in the O($^1$D) state\cite{vanDishoeck_1984}, we infer an amount of OH photodissociated per unit of time of 7$\times$10$^{10}$~cm$^{-2}$~s$^{-1}$; a value close to the formation rate $R$ via O+H$_2$ given above.

\textbf{What are the implications for the trail of water to terrestrial planets and notably to Earth?}
The enhanced D/H ratio in standard mean ocean water (SMOW)\cite{Hagemann_1970} of $1.5 \times 10^{-4}$ relative to the bulk interstellar elemental D/H ratio ($ \simeq 2 \times 10^{-5}$)\cite{Prodanovic_2010} indicates that a fraction of terrestrial water formed under cold conditions, likely at the surface of interstellar grains\cite{Cleeves_2014}. This anomaly would indeed reflect the effects of chemistry at low temperatures where the small zero-point energy difference between D- and H-bearing species can create large deuterium fractionations\cite{Tielens_1983,Ceccarelli_2014}. However, the observed D/H ratio in protostars\cite{Persson_2014,Jensen_2021,Tobin_2023} – tracing the inherited water content – is higher than the D/H ratio in SMOW ($0.3-1 \times 10^{-3}$ versus $1.5 \times 10^{-4}$, respectively). Hence, chemical processing could have occurred in warm gas, reducing the deuterium fractionation because the formation rates of HDO and H$_2$O are similar at high temperatures. In the d203-506 system, H$_2$O is efficiently destroyed and reformed via the warm gas-phase formation route $\text{O}  \rightleftarrows \text{OH} \rightleftarrows \text{H}_2\text{O}$. This can significantly decrease the water D/H ratio inherited from the cold phase to interstellar elemental values ($\simeq 2 \times 10^{-5}$)\cite{Prodanovic_2010}. The resulting HDO/H$_2$O ratio could be even lower if selective photodissociation of dihydrogen leads to subsolar values of HD/H$_2$\cite{LePetit_2002}. It is however unclear if water unveiled in the d203-506 system is incorporated into comets and asteroids since the observed OH might rather trace unbounded gas heated by the external FUV radiation. The latter is further supported by models of photoevaporative winds with similar conditions as d203-506, which predicts photodissociated OH (and consequently H$_2$O) to be primarily present in the unbounded gas\cite{Ballabio_2023}.

Nevertheless, the d203-506 system constitutes a unique interstellar laboratory as the same chemical processes will occur in the atmosphere of planet-forming disks even if the disk is solely irradiated by the host star. The FUV fields generated by accreting stars at the disk surface range from $G_0=$10$^3$ to 10$^6$ between 1 and 10~au, depending on the dust properties and shielding by the gas\cite{Bethell_2009,Bosman_2022}; a range that is comparable to that of the gas probed by our observations. 
The local density inferred from OH mid-IR lines (see Method) is somewhat lower compared with gas densities expected in disk upper layers ($0.2-1.2 \times 10^7$~cm$^{-3}$ versus $10^8-10^{12}$~cm$^{-3}$) and the temperature is higher than that found for water in planet-forming disks (1000~K versus 400-600~K)\cite{Pontoppidan_2014}. Still, the warm gas-phase formation route is already efficient at $T>400~$K and the gas density only affects the relative OH/H$_2$O ratio in the irradiated layers\cite{Zannese_2022}. 

We expect mid- and near-IR emission of OH to be detected by \textit{JWST} in a wide range of astrophysical environments (e.g., shocks, photodissociation regions at the edge of molecular clouds, comets). Our new analysis method of OH lines will provide vital constraints on the physics and chemistry of the gas in such environments. This work also constitutes a benchmark illustration of the potential of excited IR lines from small molecular species, which can be transposed to other species like CO, CH$^+$\cite{Neufeld_2021}, or CH$_3^+$. In this context, new quantum dynamical calculations and laboratory experiments are essential to exploit the full potential of \textit{JWST}.

%%%%%%%%%%%%%%%%%%%%%%%%%%%%%
%%%% Main text figures %%%%%%
%%%%%%%%%%%%%%%%%%%%%%%%%%%%%
\clearpage

\begin{table}[!h]
  \begin{center}
  \caption{Parameters derived from the analysis of OH and H$_2$ lines}
  
\label{Table:parameters}
\begin{tabular}{|l|l|c|}
   \hline

Diagnostics & Quantity & Measured value \\[0.5ex]
   \hline
   \hline
 H$_2$ lines  &$N$(H$_2$)$^a$ &  $(9.9 \pm 0.9) \times 10^{19}$ cm$^{-2}$ \\[0.5ex]
&$T^a$  &  950 $\pm$ 17~K \\[0.5ex]
 \hline
OH mid-IR lines  & $\Phi$ (rate  H$_2$O+ $h \nu$)$^{b}$ &   $(1.1 \pm 0.2) \times 10^{10}$   cm$^{-2}$ s$^{-1}$ \\[0.5ex]
&$N$(H$_2$O)     &   1 $\times 10^{15} $ -  $5 \times 10^{15}$~cm$^{-2}$              \\[0.5ex]
 & $x$(H$_2$O)$^{c}$          &   $5\times 10^{-6} $ - $2.5  \times 10^{-5} $      \\[0.5ex]
 \hline
OH near-IR lines  & $R$ (rate  O+H$_2$)$^{b,d}$       &  $(1.2 \pm 0.3) \times 10^{11}$ cm$^{-2}$ s$^{-1}$   \\[0.5ex]

 & $n_\text{H}$$^{d}$      &  $ (1.4 \pm 0.4) \times 10^7$ cm$^{-3}$       \\[0.5ex]
   \hline
\end{tabular}
\end{center}
{Note a: the uncertainties of these values do not take into account calibration effects. They could be increased by a factor of 2-3 if the uncertainty associated with calibration effects is equal to 20\%.\\
Note b: number of OH molecules formed per unit area and time via either formation route (in cm$^{-2}$ s$^{-1}$). \\
Note c: the abundance with respect to the total number of hydrogen atoms is calculated using $N(H_2)$ inferred from H$_2$ lines. \\
Note d: the inferred value would be reduced if inelastic collisional excitation were to play a significant role. The lower limit expected for these values are $R$ = 1.1$\times$10$^{10}$ cm$^{-2}$ s$^{-1}$ and $n_{\rm H}$ = 1.3$\times$10$^6$ cm$^{-3}$.}
\end{table}

\clearpage

\section*{Correspondence} Correspondence and requests for materials
should be addressed to Marion Zannese,
marion.zannese@universite-paris-saclay.fr.

\section*{Acknowledgements}

We would like to thank F. Lique for our fruitful discussion on collisions between OH and H which helped us discuss this matter. This work is based [in part] on observations made with the NASA/ESA/CSA James Webb Space Telescope. The data were obtained from the Mikulski Archive for Space Telescopes at the Space Telescope Science Institute, which is operated by the Association of Universities for Research in Astronomy, Inc., under NASA contract NAS 5-03127 for \textit{JWST}. These observations are associated with program ERS1288. D.V.D.P. acknowledges support for program \#1288 provided by NASA through a grant from the Space Telescope Science Institute
This work was supported by CNES with funds focused on \textit{JWST}. This research has been supported by the Programme National Physique et Chimie du Milieu Interstellaire (PCMI) of CNRS/INSU with INC/INP co-funded by CEA and CNES. J.R.G. thanks the Spanish MCINN for funding support under grant  \mbox{PID2019-106110GB-I00}. A.Z. acknowledges funding by Spanish MCINN under grants No. PID2019-107115GB-C21 and PID2021-122549NB-C21.
E.vD. acknowledges support from A-ERC grant N° 101019751 (MOLDISK). E.V., M.M., L.G.S. and F.J.A. acknowledge funding by Spanish MCINN  under grant  No. PID2021-122839NB-I00. 
L.G.S., P.G.J. and A.V. acknowledge funding by Spanish MCINN under grant No. PID2020-113147GA-I00. 
A.V. also acknowledges Grant No. EDU/1508/2020 (Junta Castilla y León and European Social Fund). This work is sponsored (in part) by the Chinese Academy of Sciences (CAS), through a grant to the CAS South America Center for Astronomy (CASSACA) in Santiago, Chile. C.B. is grateful for an appointment at NASA Ames Research Center through the San Jos\'e State University Research Foundation (80NSSC22M0107) and acknowledges support from the Internal Scientist Funding Model (ISFM) Laboratory Astrophysics Directed Work Package at NASA Ames. A.F. is grateful to Spanish MICIN for funding under grant PID2019-106235GB-I00. Work by Y.O. and M.R. is carried out within the Collaborative Research Centre 956, sub-project C1, funded by the Deutsche Forschungsgemeinschaft (DFG) – project ID 184018867.
E.P. and J.C. acknowledge support from the University of Western Ontario, the Institute for Earth and Space Exploration, the Canadian Space Agency, and the Natural Sciences and Engineering Research Council of Canada. T.O. acknowledges support from JSPS Bilateral Program, Grant Number 120219939. This research made use of pdrtpy, The PhotoDissociation Region Toolbox, an open-source PDR model and data analysis package
\cite{Pound_2023,Pound_2008,Kaufman_2006,Pound_2011}.

\section*{Author contributions} 

M.Z., B.T., E.H., and A.Z. wrote the manuscript with significant input from J.R.G., E.vD., A.T., E.B. and J.H. M.Z., B.T. did the line analysis with support from A.Z. and J.H.B. using the quantum dynamical calculations of M.C.vH., A.V., P.G.J., M.M., E.V., F.J.A. and L.G.S.  I.S., A.C., R.C., A.S., B.T., F.A., D.V.P. reduced the data.
E.H., E.P., and O.B. planned and co-led the ERS PDRs4All program. B.T., J.R.G., A.T., B.T., E.D., A.A., F.A., E.B., J.B.S., C.B., E.B., J.C., A.C., R.C., D.D., M.E., A.F., K.D.G., L.I., C.J., O.K., B.K., O.L., D.L., R.L.G., A.M., R.M., Y.O., T.O., S.P., M.W.P., M.R., M.R., B.S., T.S., I.S., A.S., D.V.D.P., S.V., M.G.W. contributed to the observing program with \textit{JWST}. All authors participated in either the development and testing of the MIRI-MRS or NIRSpec instruments and their data reduction, in the discussion of the results, and/or commented on the manuscript.

\section*{Methods}
\label{methods}

\paragraph{Observations and data reduction} Observations with MIRI-MRS in the integral field unit (IFU) mode were obtained as part of the Early Research Science (ERS) program PDRs4All: Radiative feedback from massive stars (ID1288, PIs: Berné, Habart, Peeters) \cite{ERS_2022}. The observations cover a 9 x 1 mosaic centered on $\alpha_{\rm J2000}$ = 05$^{\rm h}$35$^{\rm min}$20.4749$^{\rm s}$, $\delta_{\rm J2000}$ = -05$^\circ$25'10.45". A 4-point dither optimized for extended sources was applied and the FASTR1 readout pattern adapted for bright sources was used. The integration time was 521.7s using 47 groups/integration and 4 integrations. The data span a wavelength range from 4.90 to 27.9 $\mu$m, have a spectral resolution R $\sim$ 1700-3700, and a spatial resolution of 0.2"-0.8". The latter corresponds to a very small spatial scale of about 100 AU at the distance of the Orion Bar (414 pc \cite{Menten_2007}). 
The MIRI-MRS data were reduced using version 1.11.1 of the \textit{JWST} pipeline\footnote{\url{https://jwst-pipeline.readthedocs.io/en/latest/}}, and \textit{JWST} Calibration Reference Data System\footnote{\url{https://jwst-crds.stsci.edu/}} (CRDS) context 1097. The stage 2 residual fringe correction was applied in addition to the standard fringe correction step. A master background subtraction was applied in stage 3 of the reduction. The 12 cubes (4 channels of 3 sub-bands each), all pointing positions combined, were stitched into a single cube (Canin et al., in prep.) (see ref.\cite{Chown_2023} for data reduction details). The mosaic was positioned to overlap with the PDRs4All NIRSpec IFU observations of the Orion Bar which is also a 9 x 1 mosaic. The NRSRAPID readout mode appropriate for bright sources, and a 4-point dither pattern were also used for NIRSpec. The on-source integration time was 257.7 per exposure with five groups per integration with one
integration per exposure. The data were reduced using the \textit{JWST} Science Calibration Pipeline (version 1.10.2.dev26+g8f690fdc) and the context jwst\_1084.pmap of the Calibration References Data System (CRDS) (see ref.\cite{Peeters_2023} on data reduction process). The MIRI-MRS and NIRSpec spectrum, in units of MJy sr$^{-1}$, were averaged in a circular aperture centered on $\alpha_{\rm J2000}$ = 5$^{\rm h}$35$^{\rm min}$20.3145$^{\rm s}$ and $\delta_{\rm J2000}$=-5$^\circ$25'05.528" with a radius of 0.15". An OFF position was averaged in a circular aperture centered on $\alpha_{\rm J2000}$ = 5$^{\rm h}$35$^{\rm min}$20.2539$^{\rm s}$ and $\delta_{\rm J2000}$=-5$^\circ$25'05.498" with a radius of 0.3".

\paragraph{Post-processing of the \textit{JWST} spectra.} In order to better visualize OH lines in Figs. \ref{fig:MIRI_spectrum} and \ref{fig:NIRSpec_spectrum}, the continuum was subtracted considering several points in the spectral regions free of line emission and lines other than that of OH were removed as shown in Extended Data Fig. \ref{fig:spectrum_processed}. In the selected MIRI-MRS spectral range (9-10.4$\mu$m), several prominent lines are present, such as HI lines 8-13 at 9.261 and 9-20 at 9.392 $\mu$m, H$_2$ 0-0 S(3) line at 9.665 $\mu$m. The lines from the background of the planet-forming disk were subtracted by fitting them in the OFF spectrum. This allows subtracting the emission of the recombination lines of HI, 8-13 at 9.261 and 9-20 at 9.392 $\mu$m. We also subtracted the H$_2$ 0-0 S(3) line by fitting it directly in the ON spectrum (see Extended Data Fig. \ref{fig:spectrum_processed}). The same processing was applied to the NIRSpec spectrum; a continuum was subtracted and HI lines, 5-11 at 2.873 $\mu$m and 5-10 3.039 $\mu$m, OI lines at 2.893 and 3.099 $\mu$m, H$_2$ lines 2-1 O(3) at 2.974 $\mu$m, 1-0 O(4) at 3.004 $\mu$m, 2-1 O(4) at 3.189 $\mu$m and two unidentified lines at 3.164 and 3.224 $\mu$m were removed by fitting them directly in the ON spectrum.

\paragraph{Planet-forming disk d203-506 characteristics.} The planet-forming disk is an almost edge-on disk seen in silhouette against the bright background. The measured radius is $R_{\rm out}$ = 98 $\pm$ 2 au and the total mass is estimated to be about 10 times the mass of Jupiter\cite{Berné_2024}. Kinematic studies with ALMA\cite{Berné_2024} point to a stellar mass of the host star to be  below 0.3 M$_{\odot}$. This environment is the archetype of a young planet-forming disk in the intermediate phase between being embedded and being unshielded in the surrounding HI gas irradiated by an external FUV field. d203-506 shows no signs of the presence of an ionization front\cite{Storzer_1998,Bally_2000,Berné_2024}, indicating that the radiation field reaching the disk is completely dominated by FUV photons (E < 13.6 eV).
The Far-UV (FUV) radiation field incident on the ionization front (IF) of the bar is $G_0= 2-7 \times 10^4$ in Habing units\cite{Habing_1968,Peeters_2023} (G$_0=1$ corresponding to a flux integrated between 91.2 and 240~nm of $1.6\times 10^{-3}$ erg cm$^{-2}$ s$^{-1}$) as derived from UV-pumped IR-fluorescent lines \cite{Marconi_1998,Peeters_2023}. d203-506 can be illuminated by the Trapezium cluster ($\Theta^1$ Ori C) but also by $\Theta^2$ Ori A \cite{Haworth_2023}. The estimation of the FUV field at the surface of the disk is uncertain but it is expected to be similar than at the IF as determined with measures by geometrical considerations (distance between the disk and both stars) and UV-pumped IR-fluorescent lines (OI and H$_2$)\cite{Berné_2024}.
Moreover, H$_2$ ro-vibrational lines highlight a bright molecular emission originating from a photevaporative wind surrounding the edge-on disk. As seen in Fig. \ref{fig:proplyd}, there is a bright emission spot visible in the H$_2$ emission in the northwestern part of the d203-506 system. This spot appears to coincide with the region of interaction between a jet and the wind. It is only visible on the side facing the Trapezium. The molecular lines (such as OH or pure rotational H$_2$ lines) between the bright spot and the more diffuse wind are similar in relative intensity, suggesting that the molecular excitation is primarily powered by the external FUV. However, shocks could impact the bright spot's chemistry because the collimated jet could compress the gas, increasing the density locally, explaining the brightness of the spot.  

In this bright spot, we detect a very active warm water chemistry. We can estimate two timescales to understand its origin. The chemical timescale for water chemistry is determined using the photodissociation rate of water $k_{\rm \Phi}$: $\tau_{\rm photo} = 1/k_{\rm \Phi}$. This estimation leads to a chemical timescale of about a day. In comparison, the dynamical timescale of the outflowing gas $\tau_{\rm dyn}$ is much longer. It can be estimated as $\tau_{\rm dyn} = d/v_{\rm adv}$, where $d$ is the distance between the inner disk and the bright spot and $v_{\rm adv}$ is the advection velocity. In the case of d203-506 $d\sim 100$ au and $\tau_{\rm dyn} \gtrsim$ 50 years.

\paragraph{Excitation diagram of H$_2$ and OH.} The integrated line intensities of OH and H$_2$ measured with NIRSpec and MIRI-MRS are reported in Extended Data Tables \ref{tab:intensity} and \ref{tab:intensity_H2}. They were derived by fitting each line in the spectra by a Gaussian function plus linear function to account for the continuum. We note that the H$_2$ line intensities toward d203-506 are also reported in ref.\cite{Berné_2023} using a slightly different aperture. The excitation diagrams of OH and H$_2$ are derived by plotting $Y\equiv \ln{\left( \frac{4 \pi I}{h \nu_0 A_{ul} g_u} \right)}$ as a function of the upper energy level $E_{up}$ of each transition, where $I$ is the integrated line intensity, and $A_{ul}$ and $\nu_0$ are the Einstein-A coefficient and the frequency of the line, respectively. An excitation temperature $T_{ex}$ and, for H$_2$ lines, a total column density $N$ can then be inferred using the canonical formula  
\begin{equation}
Y= \ln\left(\frac{N}{Q(T_{ex})}\right) -E_u/k_B T_{ex},
\label{eq:excit_diagram}
\end{equation}
where $Q(T_{ex})$ is the partition function.

The excitation diagram of OH lines is presented in Extended Data Fig. \ref{fig:excitation_diagram} and analyzed in the main text. The excitation diagram of H$_2$ shown in Extended Data Fig. \ref{fig:diag_rot_h2} was made using the \texttt{PhotoDissociation Region Toolbox Python module}\footnote{\url{https://github.com/mpound/pdrtpy}} (pdrtpy) \cite{Pound_2023,Pound_2008,Kaufman_2006,Pound_2011}. The excitation of the ro-vibrational lines up to energy levels of $\simeq 8,000~$K can be described by a single excitation temperature of $\simeq~1,000~$K. This corresponds to the average gas temperature of the warm H$_2$ layer where OH and H$_2$O are efficiently formed and photodestroyed. We further infer from Eq. (\ref{eq:excit_diagram}) a column density of $N(H_2) = 9.9\times 10^{19}~$cm~$^{-2}$.

\paragraph{The OH model.}
Our OH model includes the ro-vibrational, $\Lambda$-doubling, and fine structure levels of OH in its ground electronic state OH($X^2 \Pi$) and in its first electronic state OH($A^2 \Sigma^+$) as provided by refs.\cite{Brooke_2016,Yousefi_2018} and compiled in ref.\cite{Tabone_2021}. The synthetic \textit{JWST} spectra of OH are calculated using the molecular excitation code \texttt{GROSBETA}\cite{Tabone_2021} and the results of quantum dynamical calculations \cite{van_Harrevelt_2000,van_Harrevelt_2001}. The excitation model is based on a single-zone approach following the formalism presented in ref.\cite{van_der_Tak_2007}. The population of OH levels is computed by considering the production of OH in various states, radiative pumping, and inelastic collisional (de-)excitation. Assuming chemical steady-state, the detailed balance equation for level i along the line of sight is given by:
\begin{equation}
	\sum_{j \neq i} P_{ji} N_j - N_i \sum_{j \neq i} P_{ij} + F \times \left( f_{i}({\rm OH}) - \frac{N_i( \rm OH)}{N( \rm OH)} \right)  = 0,
	\label{eq:statistical-eq}
\end{equation} 
where $N($OH$)$ is the total column density of OH and $N_i$ $[\text{cm}^{-2}]$ are the column densities in level $i$. $P_{ij}$ are the radiative and inelastic collisional transition probabilities. The inelastic collisional (de-)excitation of OH with He and H$_2$ was determined using inelastic collisional rate coefficients from ref.\cite{Klos2007,Offer1994} and have been extrapolated to include collisional transitions between higher rotational levels of OH such as done in ref. \cite{Tabone_2021}. In this work, the column density of OH is fixed at $N($OH$) = 2 \times 10^{15}~$cm$^{-2}$ and the density at $n_{\text{H}} = 10^{7}~$cm$^{-3}$. We further adopt a local radiation field composed of a diluted black-body at $40,000$~K to model the UV radiation field emitted by the Trapezium stars, an infrared radiation field corresponding to that detected by NIRSpec and MIRI-MRS, and a diluted black-body $50~$K to account for the far-IR and mm emission of the dust. We stress that because inelastic collisional de-excitation and radiative pumping of the OH levels probed by \textit{JWST} is orders of magnitude weaker than radiative decay, the exact values of $N($OH$)$, $n_{\text{H}}$, and local radiation field, are of little importance.

In Eq. (\ref{eq:statistical-eq}), $F$ is the formation rate, measured in cm$^{-2}$~s$^{-1}$, that produces OH in specific states i with a probability denoted as $f_i$ and we assume that the probability of destroying OH from a state i is equal to the proportion of OH in this state $N_i/N({\rm OH})$. We assume that the other formation and destruction processes do not impact the level population. 
For the sake of clarity, we compute synthetic spectra including either H$_2$O photodissociation or chemical formation-pumping. This approach is valid because these two processes excite different energy levels. Thus, we define two different formation rates: the formation rate via the photodissociation of water denoted as $\Phi$ and the formation rate via O+H$_2$ denoted as $R$. $\Phi$ is also the column density of H$_2$O photodissociated per unit of time \cite{Tabone_2021}:
\begin{equation}
        \Phi \equiv \int_{z} k_{\phi} n_{\rm H_2O} dz,
        \label{eq:phiB}
\end{equation}
where $n_{\rm H_2O}$ is the local number density of H$_2$O ([cm$^{-3}$]) and $k_{\phi}$ is the photodissociation rate of H$_2$O forming OH ([s$^{-1}$]), which depends on the strength and shape of the local FUV radiation field \cite{Heays_2017}. 
Similarly, $R$ is also the column density of OH formed per unit of time via O+H$_2$.
\begin{equation}
        R \equiv \int_{z} \sum_i \sum_j  k_{j \rightarrow i}(T) x_j({\rm H_2}) n_{\rm H_2} n_{\rm O} dz = \int_{z} k  n_{\rm H_2} n_{\rm O} dz ,
        \label{eq:R}
\end{equation}
where $n_{\rm O}$ and $n_{\rm H_2}$ are the number densities of atomic oxygen and molecular hydrogen, $x_j({\rm H_2})$ is the level population of H$_2$, and $k_{j \rightarrow i}$ is the state-to-state rate coefficient of the reaction ([cm$^{3}$~s$^{-1}$]).

The distributions of nascent OH $f_i$ stem from the results of quantum dynamical calculations and are shown in Extended Data Fig. \ref{fig:state_distrib}. For H$_2$O photodissociation, we use the rotational and vibrational state distribution of OH produced by H$_2$O photodissociation via both its $\tilde{A}$ and $\tilde{B}$ electronic states calculated by ref. \cite{van_Harrevelt_2000} and ref.\cite{van_Harrevelt_2001} and compiled by ref. \cite{Tabone_2021}. Since these pioneering calculations did not include the spin-orbit coupling nor the $\Lambda$-doubling, we include the results from ref. \cite{Zhou_2015} by assuming that photodissociation via the $\tilde{B}$ state leads to OH in symmetric states, with equal distribution between the two spin-orbit manifold (see Extended Data Fig. \ref{fig:energy_levels} for the spectroscopy of OH). The resulting state distribution of OH is calculated by integrating the state-specific cross-section over a UV field represented by a black-body at $T_{eff} = 40,000~$K (see ref.\cite{Tabone_2021}). In the absence of systematic quantum dynamical calculations for short wavelength photons, we further neglect the impact of H$_2$O photodissociation shortward of $114~$nm on the excitation of OH.

For formation-pumping of OH, we extracted the distribution of nascent OH from the state-to-state rate coefficients computed by Veselinova et al. in ref. \cite{Veselinova_2021}. Because the distribution of OH depends on both the kinetic temperature and the population of H$_2$, the distributions are computed as
\begin{equation}
    f_i({\rm OH}) =  \sum_j k_{j \rightarrow i}(T) x_j({\rm H_2}) / \sum_i \sum_j k_{j \rightarrow i}(T) x_j({\rm H_2}),
\end{equation}
where $x_j( \rm H_2)$ is the level population of H$_2$ and $k_{j \rightarrow i}(T)$ are the state-to-state rate coefficients at temperature $T$. The values of $x_j(\rm H_2)$ are inferred from the excitation diagram of H$_2$, which probes the majority of H$_2$ levels. For the H$_2(v=0, J=0)$, H$_2(v=0, J=1)$ and H$_2(v=0,J=2)$ levels, which are not probed by \textit{JWST}, we extrapolated their populations assuming local thermodynamical equilibrium. Interestingly, we find that the resulting distribution of OH differs very little from a distribution assuming a Boltzmann distribution of H$_2$ levels at $T \simeq 1,000$~K. In fact, the formation rate of OH($v=1$) is driven by the rotationally excited levels of H$_2$ ($v$=0, $J \simeq 4-9$) that are much more populated than the vibrational levels and are reactive enough to produce vibrationally excited OH by reacting with O atoms.

\paragraph{Conversion between line intensity and formation rate}
In the d203-506 system, we find that the highly excited rotational and vibrational OH levels are primarily populated by formation pumping (the reaction O+H$_2$ or the photodissociation of water) followed by radiative decay, with negligible impact of radiative pumping and inelastic collisions. Furthermore, the ro-vibrational and rotationally excited lines are found to be optically thin.
In the optically thin regime, the integrated line intensity is proportional to the column density of OH in the upper energy state:
\begin{equation}
	I_{i \rightarrow j} = \frac{ h \nu_{i,j} A_{i,j}}{4 \pi} N_i .
	\label{eq:intensity_optically_thin_n}
\end{equation} 
Assuming that excited OH levels are only populated by formation pumping and radiative decay and that the fraction of OH in rotationally and ro-vibrationally excited states constitutes
a negligible fraction of the total population of OH, the detailed balance equation (\ref{eq:statistical-eq}) simply writes:
\begin{equation}
	\sum_{j, j \neq i} M_{ji} N_j + F  \times f_{i}   = 0 ~~~\text{with}~~~\left\{\stackrel{M_{ji} = A_{ji}~~\text{if}~~i \neq j}{M_{ji} = - \sum_k A_{ik}~~\text{if}~~i=j }\right.
	\label{eq:statistical-eq-cascade}
\end{equation} 
This system of linear equations on $N_i$ shows that $N_i$, and therefore the intensity $I_{ij}$, are simply proportional to $F$, the formation rate of OH. Following ref.\cite{Tabone_2021}, we can therefore define the intensity as 
\begin{equation}
	I_{ij} = \frac{ h \nu_{i,j}}{4 \pi}  \tilde{I}_{ij} F,
	\label{eq:intensity_optically_thin_f}
\end{equation} 
where $\tilde{I}_{ij}$ is the dimensionless proportionality factor. It corresponds to the probability that a nascent OH eventually cascades via the radiative transition $i\rightarrow j$.

$\tilde{I}_{ij}$ depends only on $f_i$, which in turn depends on the shape of the radiation field when dealing with excitation by H$_2$O photodissociation, and the gas temperature and the population of H$_2$ when dealing with chemical formation pumping. For water photodissociation, $\tilde{I}_{ij}$ is provided in Appendix D of ref.\cite{Tabone_2021} for various shapes of the UV radiation field. For the chemical pumping formation via O+H$_2 \rightarrow$ OH+H, the proportionality factor computed at $T= 1,000~$K summed over all the four component of the $v=1 \rightarrow 0$, $N=4 \rightarrow 5$ ro-vibrational lines is $\tilde{I} =3 \times 10^{-3}$.

\paragraph{Estimation of column density of warm water and local gas density}

Our knowledge of the evolution of matter in space, from diffuse clouds to planets, depends on our ability to assess the local physical and chemical properties. Thanks to the derived value of the OH formation rates via the photodissociation of water $\Phi$ and via O+H$_2$ $R$, one can derive astrophysical quantities such as the column density of warm water $N$(H$_2$O) and the local gas density. Assuming an homogeneous medium  Eq. (\ref{eq:phiB}) can be rewritten as
\begin{equation}
\label{eq:NH2O}
    N({\rm H_2O})=\frac{\Phi}{k_{\phi}},
\end{equation}
where the photodissociation rate $k_{\phi}$ depends on the FUV radiation field. Based on the OH mid-IR lines, we estimate $\Phi$=1.1 $\times$ 10$^{10}$ cm$^{-2}$ s$^{-1}$ (see main text). We can obtain an estimate of the photodissociation rate of H$_2$O based on the local FUV field intensity. In the d203-506 system, the exposed reservoir of water likely lies at low visual extinction. Here, 
we assume that the FUV field at the H$^0$/H$_2$ transition is extincted by as much as a factor 5 such that $k_{\phi} \simeq 1 \times 10^{-5}-2\times 10^{-6}$~s$^{-1}$ in the bright spot, corresponding to a visual extinction of $A_V \lesssim 1$. 
This translates to $N$(H$_2$O)$\simeq$ $1 \times 10^{15} - 5 \times 10^{15}$   cm$^{-2}$.

Similarly, assuming a homogeneous medium, the local density can be estimated from the inferred value of $R$ from Eq. (\ref{eq:R}) as:
\begin{equation}
\label{eq:nH}
   n_{\rm H} = \frac{R}{k(T) N({\rm H_2}) x({\rm O})},
\end{equation}
where $k(T)$ is the rate coefficient of the reaction, $N(\rm H_2)$ is the column density of H$_2$, $x(\rm O)$ is the oxygen abundance, and $n_{\rm H}$ is the total number density of hydrogen atoms. For d203-506, we found a column density of warm excited H$_2$ of $N(\rm H_2) = 9.9 \times 10^{19}~$cm$^{-2}$ and $T \simeq 1,000~K$ from the rotational and ro-vibrational lines (see Extended Data Fig. \ref{fig:diag_rot_h2}). From the estimation of temperature and excitation of H$_2$, we can derive the coefficient rate $k=2.8 \times 10^{-13}$ cm$^3$ s$^{-1}$. Using the estimate of $R$ = 1.2$\times$10$^{11}$ cm$^{-1}$ s$^{-1}$ from the OH near-IR lines, we derive $n_{\rm H}$ = 1.4$\times$10$^7$ cm$^{-3}$.

\paragraph{Ro-vibrational excitation of OH}
In this work, we provide evidence that the near-IR OH lines are excited by chemical pumping via O+H$_2$ $\rightarrow$ OH+H. In this section, we review alternative processes that can lead to vibrationally excited OH in the ground electronic state.

Water photodissociation via the $\tilde{A}$ electronic state by long wavelength photons produces rotationally cold, but vibrationally hot OH. Our OH model does include this process using the OH distribution computed as described in ref.\cite{van_Harrevelt_2001} while using the improved H$_2$O $\tilde{A}$ state potential energy surface of ref.\cite{Harrevelt_2000}.
The resulting OH emission spectrum when considering only water photodissociation via both the $\tilde{A}$ and $\tilde{B}$ states is shown in Extended Data Fig. \ref{fig:nearIR_alternative}. The overall shape of the modeled near-IR spectrum is very different from that observed. The ro-vibrational line within the $v$=1-0 band exhibits a much lower rotational temperature, and the $v$=2-1 band is predicted to be strong.
Adopting the value of $\Phi$ that fits the mid-IR lines of OH, we also find that the ro-vibrational lines typically a factor 10-20 weaker than the observed near-IR lines.

UV radiative pumping of OH via the A-X band can also produce vibrationally excited OH in the ground electronic state. Our excitation model does take into account this process since the OH levels extend to the non-dissociative OH($A^2 \Sigma^+$) electronic state. Discarding OH chemical pumping in the excitation model, we find that the OH near-IR spectrum is about two to three order of magnitude weaker than observed, and exhibits a too-cold rotational temperature within the $v=1$ state. We note that for UV pumping, the intensity of the lines depends on the assumed column density of OH, which is not directly probed by our observations. However, the adopted column density of $N(\rm OH) =  2 \times 10^{15}$~cm$^{-2}$ is a good first-order estimate, since it corresponds to an OH abundance of $x($OH$) \simeq 10^{-5}$, a value that is in line with the results from physicochemical models\cite{Zannese_2022}. 

Little is known about the inelastic collisional cross sections connecting the $v=0$ and the $v=1$ states of OH. Owing to the repulsive barrier between OH and H$_2$, and OH and He, collisions with H$_2$ and He are negligible compared to those with atomic hydrogen, which is likely abundant in the OH emitting layer. Collision with the latter is thus the only process that can contribute to the excitation of OH($v=1$). In fact, ref.\cite{Atahan_2006} computed the inelastic de-excitation collisional rate coefficients for OH(v=1)+H $\rightarrow$ OH(v=0) + H below a kinetic temperature of $T=300~$K and assuming a thermal distribution of OH levels. In the d203-506 system, the temperature is much higher, about $T\simeq 1,000~$K, and the rotational levels of OH even in the $v=0$ are predicted to be subthermally populated due to the low gas density ($n_{\text{H}} \simeq 10^7~$cm$^{-3}$). Extrapolating the result of ref.\cite{Atahan_2006} to the conditions in the OH emitting gas remains uncertain. Still, for a temperature of $T \simeq 1,000~$K, we estimate a rate of the order of $k_{\rm coll} \simeq 5 \times 10^{-13}$cm$^{3}$~s$^{-1}$. The ratio between the OH($v$=1) formed by chemical formation-pumping and by inelastic collisions is about $x( {\rm OH}) k_{\rm coll} / x($O$) k_{\rm chem, v=1}$, where $k_{\rm chem, v=1}$ is the formation rate O+H$_2$ leading to OH($v=1$). 
We conclude that chemical pumping dominates over inelastic collisional excitation for an abundance ratio of about $x($O$)/x($OH$) \gtrsim 10$. The calculation of inelastic collisional rate coefficients is warranted to evaluate the contribution of inelastic collisions in the emission of OH. If inelastic collisions were to contribute to the near-IR emission of OH, the inferred chemical formation rate $R$ would be lower. However, we stress that the formation rate of O+H$_2 \rightarrow$ OH+H is close to the destruction rate of OH through photodissociation, as inferred from [OI] emission. This provides further evidence that the bulk part of OH emission is produced via chemical formation and not by inelastic collisional excitation.

\paragraph{Quantum dynamical calculations for O+H$_2$.} This section summarizes the \textit{ab initio} quantum determination of the \mbox{state-to-state} rate coefficients of the reaction
\mbox{O\,($^3P$)\,+\,H$_2$\,($v$,$j$)\,$\rightarrow$ OH\,($v'$,$j'$)\,+\,H}
used in our OH excitation and chemical models, see also ref.\cite{Veselinova_2021}.
Time-independent quantum mechanical (QM) scattering calculations were carried out applying the coupled-channel hyperspherical method implemented in the ABC code \cite{SCM:CPC00} yielding the usual S matrix, from which the state-to-state cross sections and rate coefficients can be computed. 
The set of potential energy surfaces (PESs) calculated by ref.\cite{zanchetjcp19} was used, which includes the PESs of the two electronic states (one of symmetry  $^3A''$ and one of symmetry $^3A'$ ) that correlate adiabatically with the electronic ground state of reactants, O($^3P$)+H$_2$, and products, OH($^2\Pi$)+H. 
Scattering calculations were carried out at 60 energies between 0.3-2.5 eV, including all partial waves ($J$) needed to reach convergence ($J_{\rm max}=62$) and all the values of the helicity ($\Omega$) up to 26. QM scattering calculations provided the cross section as a function of the rotational quantum number assuming that OH is a closed-shell molecule. To calculate the nascent   rotational  population of OH in terms of the total angular momentum excluding spin, N, we plotted  the calculated  cross sections calculated as a function of the rotational energy, and assigned to the N state the cross section corresponding to the actual energy of that state. Calculations do not distinguish between the two spin-orbit manifolds, and the procedure of ref. \cite{jambrinanc16} was used to distribute the population of the product's rotational states into the two $\Lambda$-doublet components.
The propagation was carried out in 300 log-derivative steps, with a maximum value of the hyperradius of 20 $a_0$. The basis included all the diatomic energy levels up to 3.25 eV.

Since reactions can occur on both  $^3A'$ and $^3A''$ electronic states,  the total rate constant of the reaction is obtained as a weighted average of the respective rates k$_{A'}$ and k$_{A''}$.
The weights are obtained by establishing the correlation between the electronic states $A'$ and $A''$ and the spin-orbit levels of oxygen, and are thus related to their relative population.
It is found that the 1$^3A''$ state correlates with three of the five components of O($^3P_{2}$), while the 1$^3A'$ state correlates with the other two components of O($^3P_{2}$) and one component of O($^3P_{1}$).
Under these considerations and assuming a thermal distribution of the spin-orbit levels of the oxygen atom, the overall thermal rate coefficients are calculated as ref.\cite{Veselinova_2021}: 
\begin{equation} \label{rate}
k(T) = \frac{3  k_{A''}(T) + \left[ 2 + \exp\big(-\frac{\Delta E_1}{T}\big)
\right] k_{A'}(T)}{5 + 3\cdot \exp\big(-\frac{\Delta E_1}{T}\big) +
\exp\big(-\frac{\Delta E_0}{T}\big)},
\end{equation}
 where $k(T)$ referrers to the desired \mbox{($v,j \rightarrow v', j'$)} state-to-state rate coefficient at a given temperature, $k_{A'}(T)$ and $k_{A''}(T)$ are the state-to-state rate coefficients on the $A'$ and $A''$ PESs, respectively.   
\mbox{$\Delta E_1$=227.708 K} and \mbox{$\Delta E_0$=326.569 K} are the energies of the oxygen atom spin-orbit levels $^3P_1$ and $^3P_0$, respectively, over the ground state $^3P_2$. We should note that the $^3P_0$ contribution and two contributions of the $^3P_1$ do not appear in the numerator; this reflects the fact that these spin-orbit states are not reactive and will not contribute to the total rate constants of OH formation.

\section*{Data Availability}

The \textit{JWST} data presented in this paper are publicly available through the MAST online archive (\url{http://mast.stsci.edu}) using the PID 1288. \\

\section*{Code Availability}

The JWST pipeline used to produce the final data products presented in this article is available at \url{https://github.com/spacetelescope/jwst}. The \texttt{GROSBETA} code used in this study is available from the corresponding author on reasonable request.

\section*{Competing Interests}
The authors declare no competing financial interests.\\ 

\bibliography{scibib.bib}

\clearpage

%\begin{appendices}

\section*{Extended Data}\label{secA1}
\renewcommand{\figurename}{Extended Data Figure} 
\setcounter{figure}{0}  
\renewcommand{\tablename}{Extended Data Table} 
\setcounter{table}{0}  

\begin{figure}
    \centering
    \includegraphics[width=\linewidth]{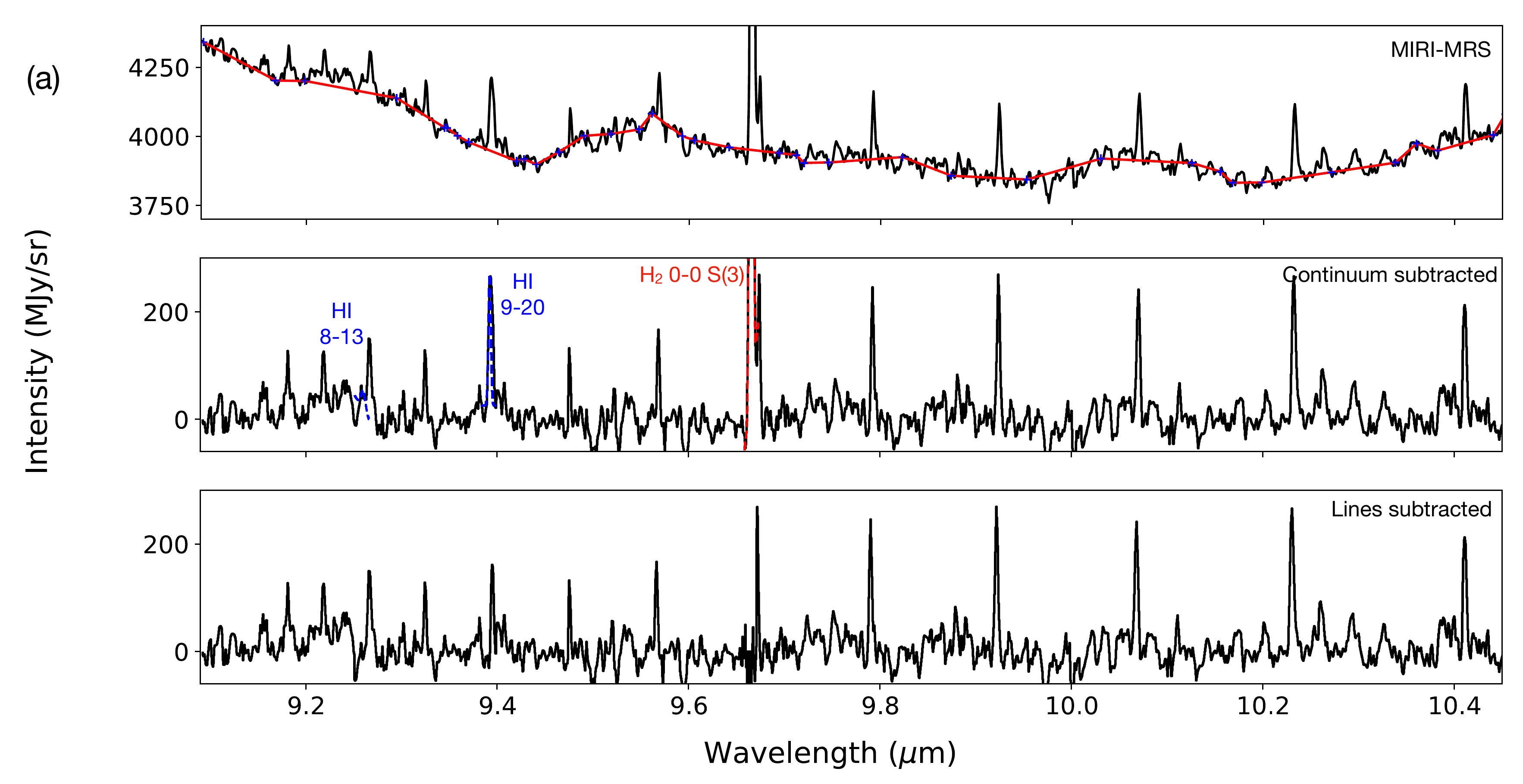}
    \includegraphics[width=\linewidth]{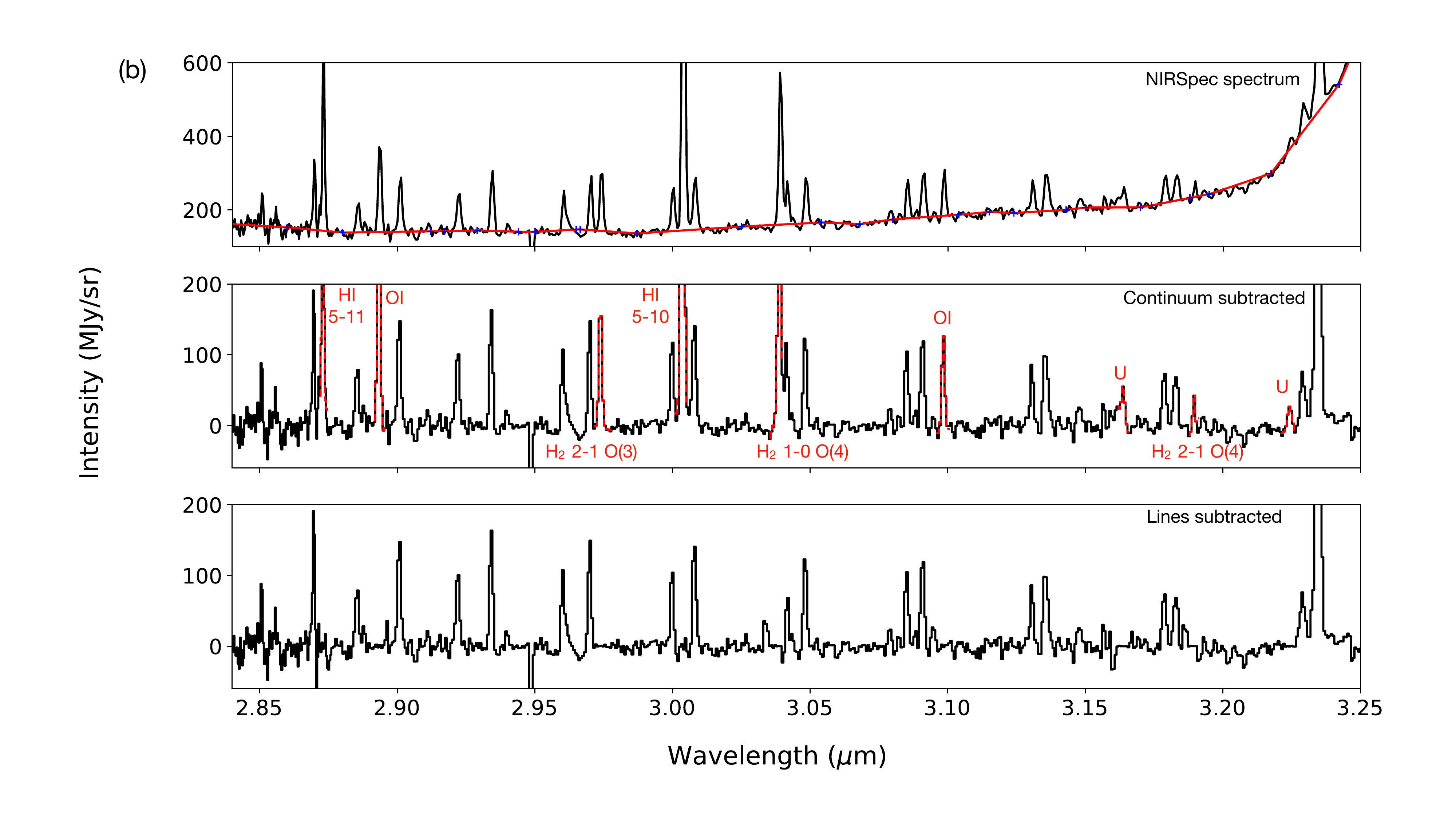}
    \caption{(a) (Top panel) Spectrum observed with MIRI-MRS. The red line is the estimated continuum. (Middle panel) Continuum subtracted spectrum. The red Gaussians are the fits to lines other than OH. The blue Gaussians are the fits to lines from the OFF position which contaminate OH lines. (Bottom panel) Processed spectrum with the dust continuum and the bright lines other than OH subtracted. (b) (Top panel) Spectrum observed with NIRSpec. The red line is the estimated continuum. (Middle panel) Continuum subtracted spectrum. The red Gaussians are fits to lines other than OH. (Bottom panel)  Processed spectrum with the dust continuum and the bright lines other than OH subtracted.}
    \label{fig:spectrum_processed}
\end{figure}

\begin{figure}
    \centering
    \includegraphics[width=0.99\linewidth]{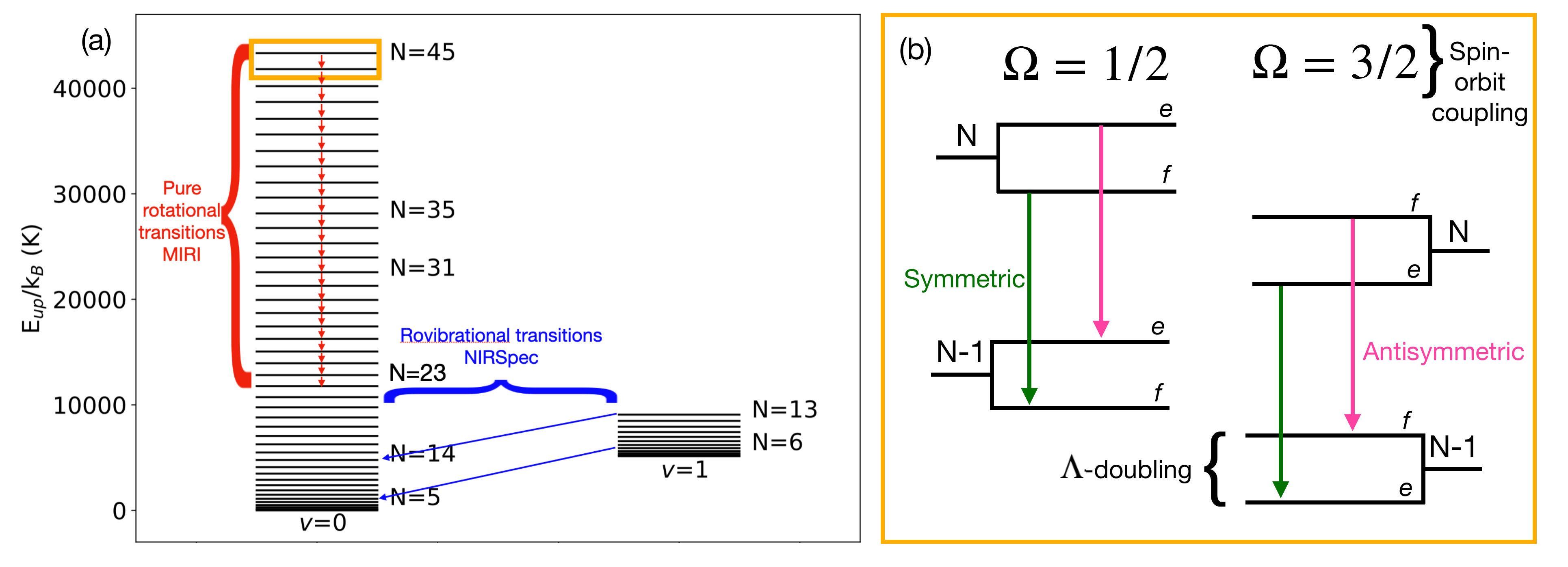}
    \caption{Schematic view of OH spectroscopy relevant to \textit{JWST}. (a) OH rotational and ro-vibrational energy levels. The red arrows are the detected pure rotational transitions with MIRI-MRS. The blue arrows are some of the detected ro-vibrational transitions with NIRSpec. (b) Zoom on the yellow box to reveal the splitting of a rotational level due to the spin-orbit coupling and the $\Lambda$-doubling. The two spin-orbit states are labeled by the $\Omega$ quantum number and the $\Lambda$-doubling states are labeled by their $\epsilon$ = \textit{e / f} spectroscopic parity. The green and pink arrows are the transitions detected in the observations forming a quadruplet. The green arrows are the transitions arising from symmetric states and the pink arrows are the transitions arising from anti-symmetric states.}
    \label{fig:energy_levels}
\end{figure}

\begin{figure}
    \centering
\includegraphics[width=\linewidth]{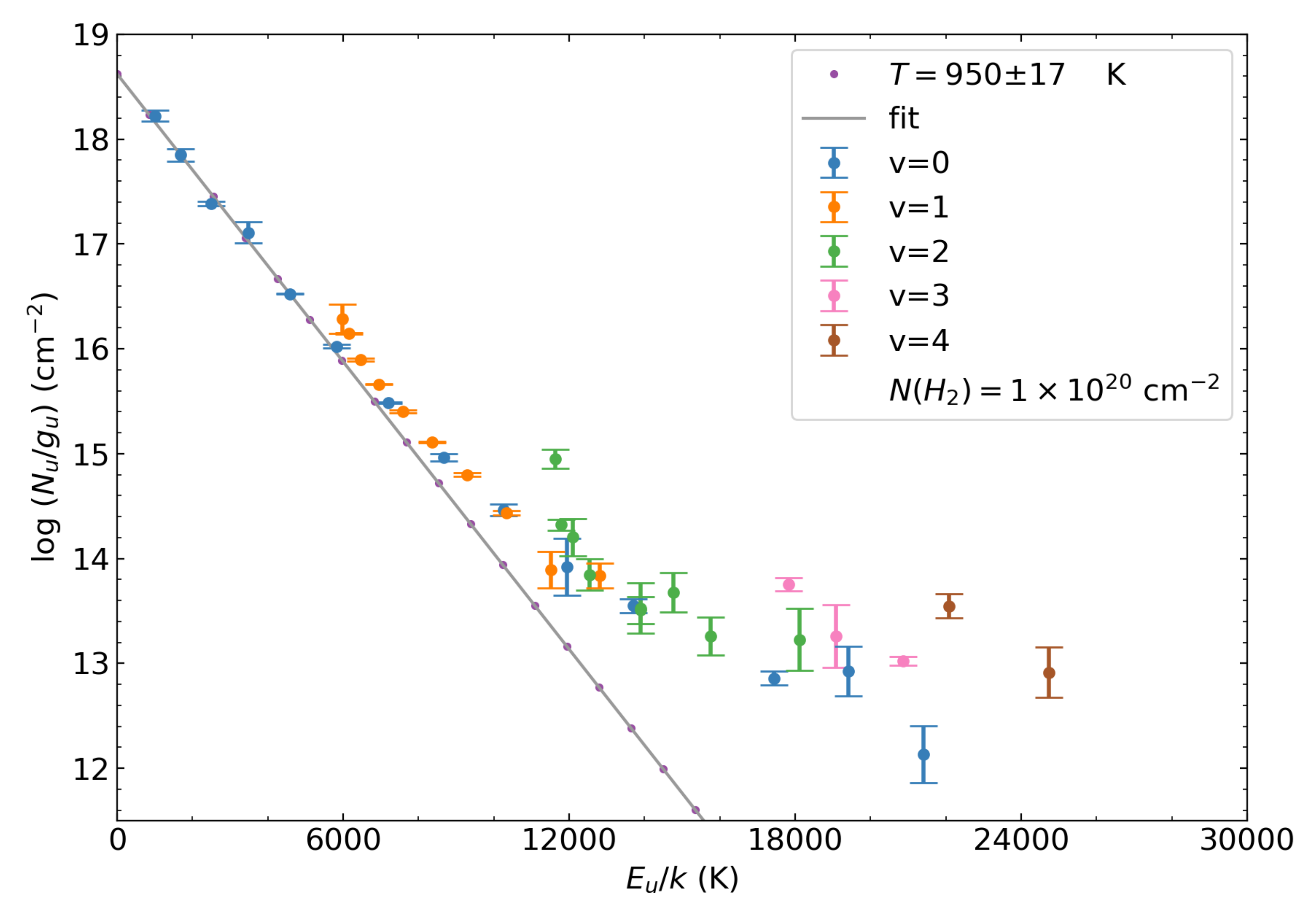}
    \caption{Excitation diagram of H$_2$ at the bright spot. The temperature fit was made on the first five pure rotational lines. The measured line fluxes are reported in Extended Data Table \ref{tab:intensity_H2}.}
    \label{fig:diag_rot_h2}
\end{figure}

\begin{figure}
    \centering
    \includegraphics[width=0.75\linewidth]{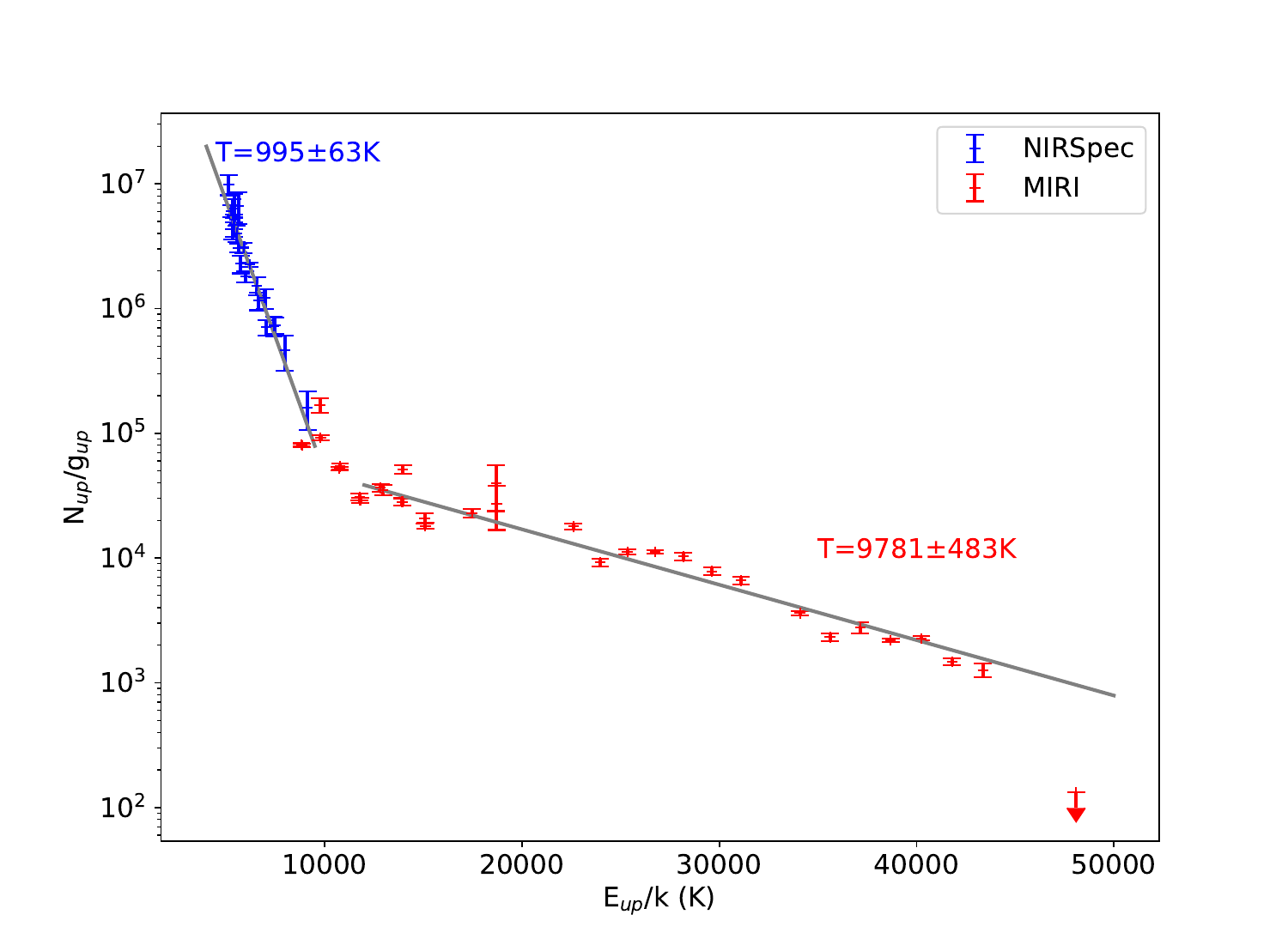}
    \caption{Excitation diagram of ro-vibrational and pure rotational lines of OH. The blue crosses are the ro-vibrational transitions detected with NIRSpec ($v$=1-0) and the red crosses are the pure rotational transitions detected with MIRI ($v$=0-0). The measured line fluxes are reported in Extended Data Table \ref{tab:intensity}.}
    \label{fig:excitation_diagram}
\end{figure}

\begin{figure}
    \centering
    \includegraphics[width=\linewidth]{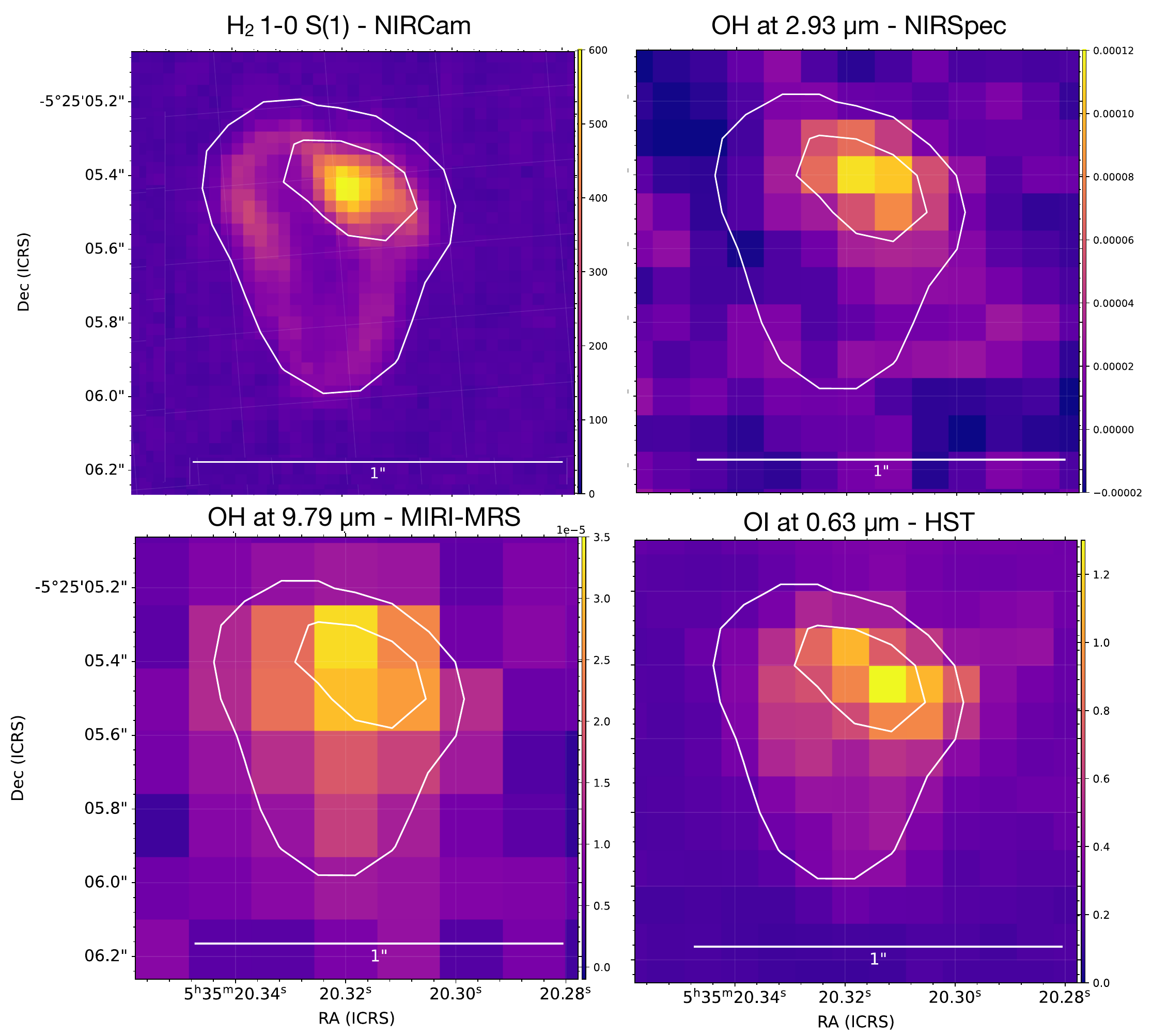}
    \caption{Spatial distribution of the emission of key lines. (Top Left) d203-506 image from the NIRCam F212N filter (in MJy sr$^{-1}$), (Top Right) integrated intensities image from NIRSpec OH line at 2.934 $\mu$m (in erg cm$^{-2}$ s$^{-1}$ sr$^{-1}$), (Bottom Left) integrated intensities image from MIRI-MRS OH line at 9.791 $\mu$m (in erg cm$^{-2}$ s$^{-1}$ sr$^{-1}$), (Bottom Right) integrated intensities image from Hubble Space Telescope (HST) OI at 0.63 $\mu$m (in counts s$^{-1}$)\cite{Bally_2000}. Contours of NIRSpec  H$_2$ 1-0 S(1) line are shown in white (at 7 $\times$ 10$^{-4}$, 2 $\times$ 10$^{-3}$ erg cm$^{-2}$ s$^{-1}$ sr$^{-1}$).}
    \label{fig:map_OH}
\end{figure}

\begin{figure}
    \centering
    \includegraphics[width=1.\linewidth]{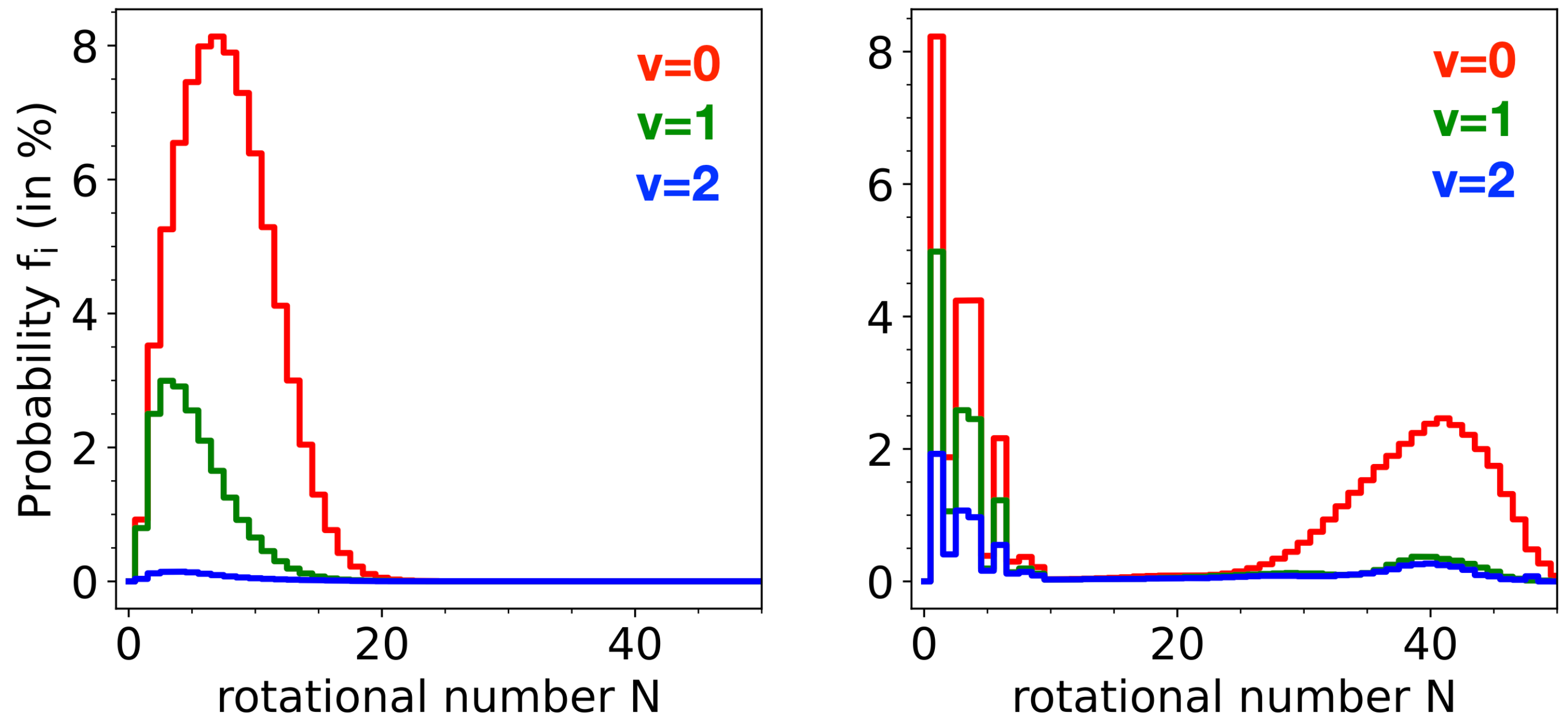}
    \caption{State distributions of nascent OH predicted by quantum dynamical calculations summed over the spin-orbit and $\Lambda-$doubling states. (Left) Distribution of nascent OH following its formation via O+H$_2$ at a temperature of $T=1,000~$K and for a H$_2$ population as inferred for d203-506. (Right) Distribution of nascent OH following H$_2$O photodissociation by an FUV field representative of the Orion Bar. Photodissociation by short wavelength photons $\lambda <144~$nm via $\tilde{B}$ electronic state of water produces rotationally hot OH with rotational quantum numbers of $N \simeq 35-45$ and photodissociation by longer wavelength photons via the $\tilde{A}$ electronic state leads to rotationally cold but vibrationally hot OH.}
    \label{fig:state_distrib}
\end{figure}

\begin{figure}
    \centering
    \includegraphics[width=1.\linewidth]{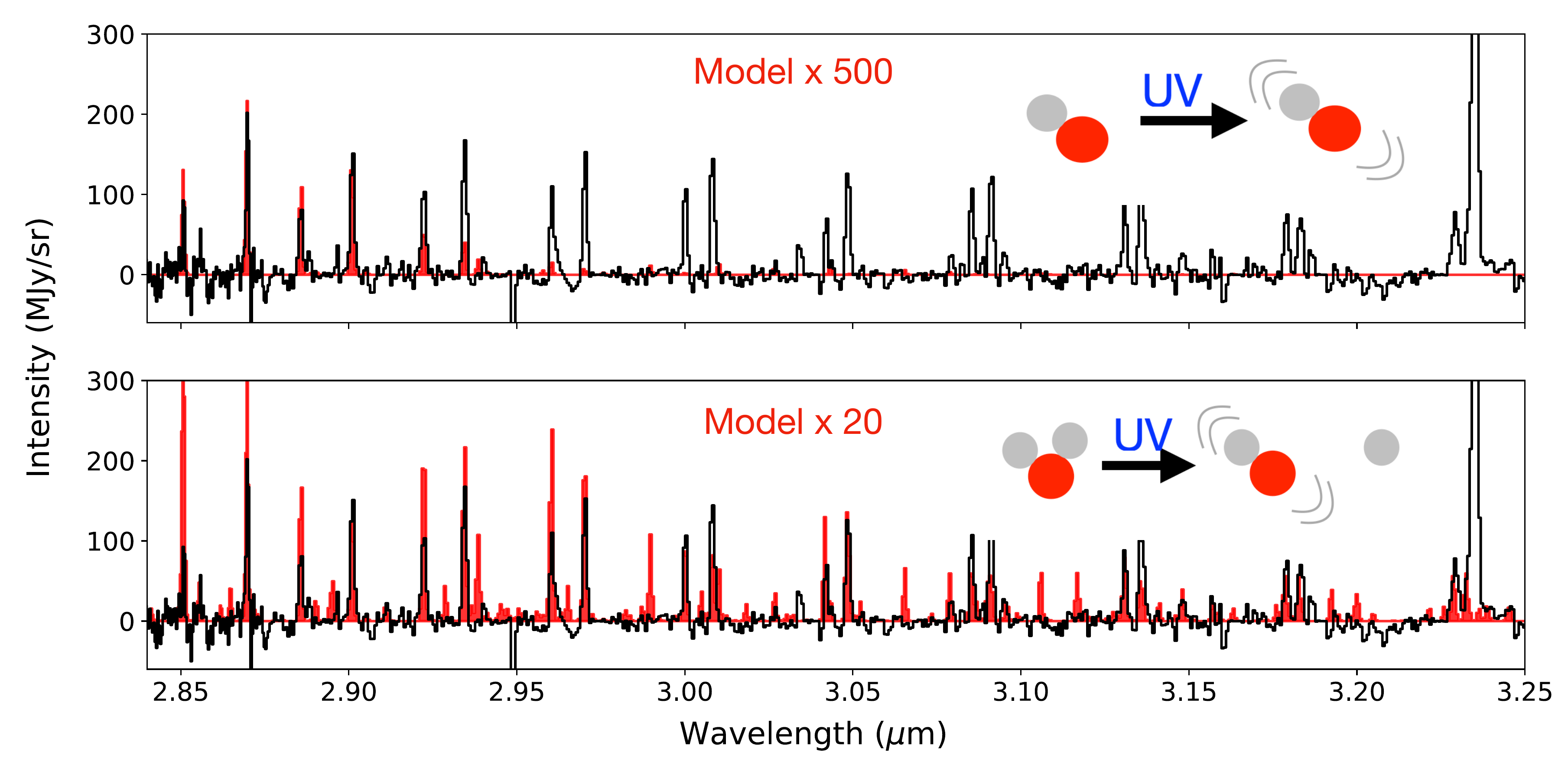}
    \caption{OH near-IR synthetic \texttt{GROSBETA} models for (Top) UV radiative pumping and (Bottom) H$_2$O photodissociation via its $\tilde{A}$ electronic state. None of these processes can account for the shape and  strength of the observed OH ro-vibrational spectrum.}
    \label{fig:nearIR_alternative}
\end{figure}

\begin{figure}
    \centering
    \includegraphics[width=\linewidth]{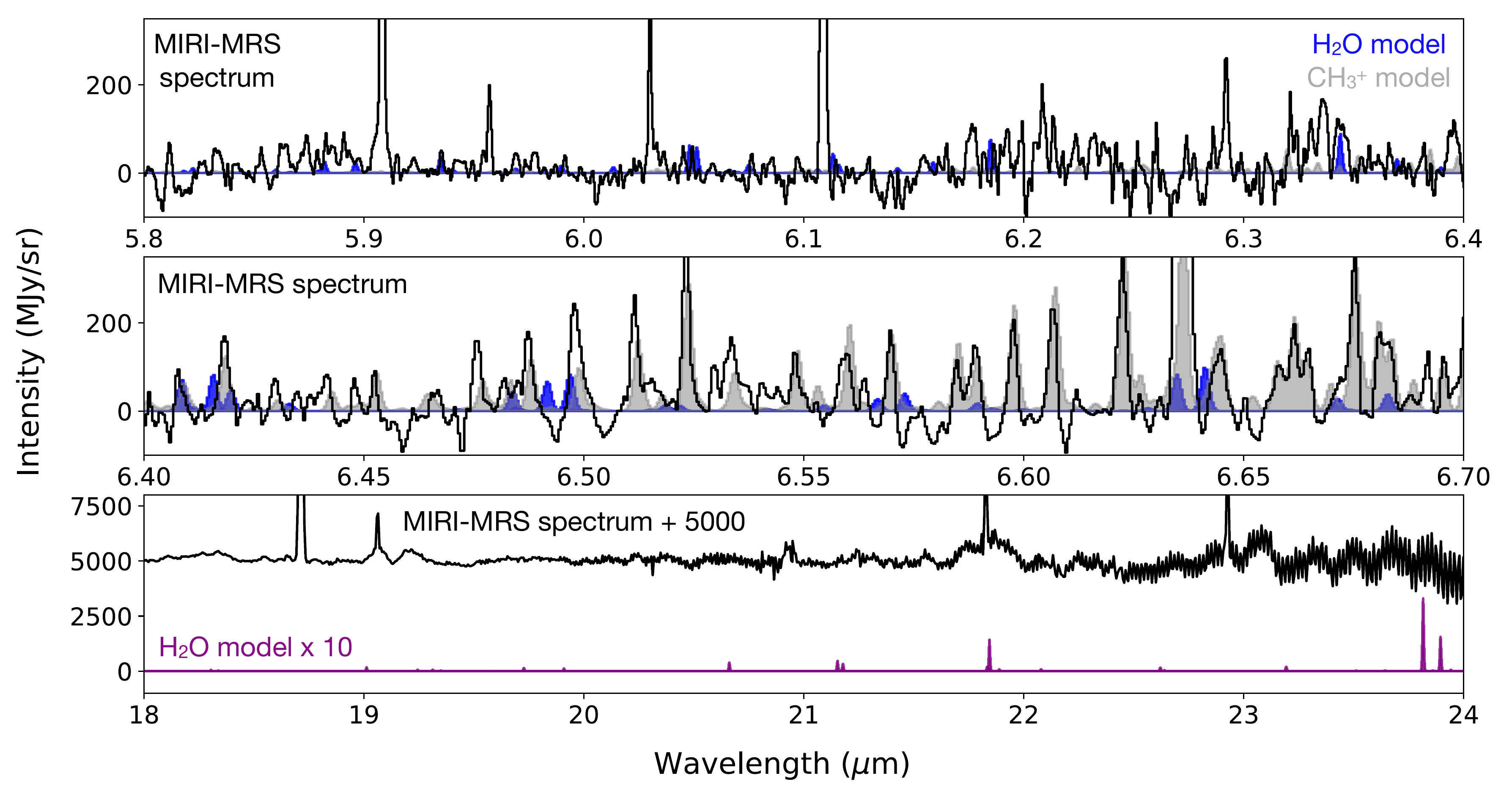}
    \caption{MIRI-MRS observations versus a synthetic \texttt{RADEX}\cite{van_der_Tak_2007} spectrum of the brightest H$_2$O lines adopting the maximum amount of unseen water of $N(H_2O) = 5 \times 10^{15}$~cm$^{-2}$ as inferred from mid-IR lines of OH, a temperature of $T$=1000K as inferred from H$_2$ lines and a density of $n_{\text{H}} = 10^{7}$~cm$^{-3}$ as inferred for near-IR lines of OH. Inelastic collisional rate coefficients are from ref.\cite{Faure_2008} available on the LAMDA database\cite{Schoier_2005}. We  also assumed an IR background inferred from NIRSpec and MIRI-MRS observations. When calculating the line intensities, we assumed that the IR continuum background interacting with the gas is not along the same line of sight as the observations are taken as described in ref.\cite{Tabone_2021}. The latter assumption provides a strict upper limit on the line strength. Owing to the low gas density, H$_2$O is subthermally excited, leading to undetectable lines. In the 5-7 $\mu$m region, the MIRI-MRS spectrum is affected by residual fringe and possible contamination by CH$_3^+$\cite{Berné_2023}. A LTE model at $T=700$K of CH$_3^+$ is overlayed in grey\cite{Changala_2023}. In the 18-24$\mu$m, the continuum increases greatly so the noise increases accordingly making the detection of H$_2$O lines in this region impossible. The same model of H$_2$O multiplied by a factor of 10 is overlayed in purple.}
    \label{fig:H2O_GROSBETA}
\end{figure}

\clearpage

\begin{table}
\footnotesize
    \centering
    \begin{tabular}{c|c|c|c}
     \multicolumn{4}{c}{MIRI-MRS} \\
        \hline 
    
        $\lambda$ ($\mu$m) & $N$ & $E_{\rm up}/k_B$ (K) & $I$ ($\times$10$^{-5}$ erg cm$^{-2}$ s$^{-1}$ sr$^{-1}$) \\
        \hline
       % 9.15  & 45 & 44,939 & 2.52 $\pm$ 0.60 \\
        9.181  & 44 & 43,343 & 0.91 $\pm$ 0.12 \\
        9.219  & 43 & 41,776 & 1.02 $\pm$ 0.07 \\
        9.266  & 42 & 40,216 & 1.50 $\pm$ 0.06 \\
        9.325  & 41 & 38,663 & 1.38 $\pm$ 0.05 \\
        9.394  & 40 & 37,120 & 1.64 $\pm$ 0.16 \\
        9.475  & 39 & 35,588 & 1.30 $\pm$ 0.09 \\
        9.568  & 38 & 34,070 & 1.89 $\pm$ 0.08 \\
        9.791  & 36 & 31,079 & 2.97 $\pm$ 0.21 \\
        9.923  & 35 & 29,609 & 3.24 $\pm$ 0.23 \\
        10.069 & 34 & 28,160 & 3.89 $\pm$ 0.30 \\
        10.232 & 33 & 26,731 & 3.86 $\pm$ 0.12 \\
        10.411 & 32 & 25,325 & 3.48 $\pm$ 0.18 \\
        10.608 & 31 & 23,943 & 2.58 $\pm$ 0.18 \\
        10.825 & 30 & 22,586 & 4.44 $\pm$ 0.27 \\
        11.610 & 27 & 18,671 & 3.38 $\pm$ 1.35 \\
        11.615 & 27 & 18,701 & 2.24 $\pm$ 0.87 \\
        11.929 & 26 & 17,447 & 3.30 $\pm$ 0.26 \\
        12.657 & 24 & 15,053  & 1.12 $\pm$ 0.10 \\
        12.665 & 24 & 15,085 & 0.92 $\pm$ 0.05 \\
        13.081 & 23 & 13,916 & 1.27 $\pm$ 0.09\\
        13.089 & 23 & 13,949   & 2.20 $\pm$ 0.18 \\
        13.550 & 22 & 12,816  & 1.36 $\pm$ 0.09 \\
        13.560 & 22 &  12,850 & 1.24 $\pm$ 0.12 \\
        14.070 & 21 & 11,754  & 0.94 $\pm$ 0.07 \\
        14.082 & 21 & 11,789  & 0.84 $\pm$ 0.04 \\
        14.648 & 20 & 10,783 & 1.29 $\pm$ 0.04 \\
        14.664 & 20 &  10,767 & 1.27 $\pm$ 0.07 \\
        15.295 & 19 & 9,749  & 3.34 $\pm$ 0.44 \\
        15.314 & 19 & 9,786  & 1.71 $\pm$ 0.08 \\
        16.021 & 18 &  8,809 & 1.26 $\pm$ 0.03 \\
        16.045 & 18 & 8,847  & 1.16 $\pm$ 0.04 \\
        \end{tabular}
       \begin{tabular}{c|c|c|c|c}
        \multicolumn{5}{c}{NIRSpec} \\
        \hline 
        
        $\lambda$ ($\mu$m)  & $\Omega$ & N & $E_{\rm up}/k_B$ (K) & $I$ ($\times$10$^{-5}$ erg cm$^{-2}$ s$^{-1}$ sr$^{-1}$) \\
        \hline
       % 2.70 & 3/2 & 4  & 5627 & 2.56 $\pm$ 0.72 \\
        2.718 & 1/2 & 3 & 5414 & 2.06 $\pm$ 0.65 \\
        2.851 & 1/2 & 1  & 5314 & 2.89 $\pm$ 1.02\\
        2.870 & 3/2 & 1  & 5134 & 6.77 $\pm$ 1.24 \\
        2.886 & 1/2 & 2  & 5401 & 4.81 $\pm$ 0.97 \\
        2.901 & 3/2 & 2  & 5250 & 7.92  $\pm$ 0.86 \\
        2.922 & 1/2 & 3  & 5541 & 6.35 $\pm$ 0.96 \\
        2.934 & 3/2 & 3  & 5415 & 10.00 $\pm$ 0.41 \\
        2.960  & 1/2 & 4 & 5735 & 4.99 $\pm$ 0.81\\
        2.970 & 3/2 & 4  & 5627 & 7.99 $\pm$ 0.61 \\
        3.000 & 1/2 & 5  & 5982 & 5.12 $\pm$ 0.55 \\
        3.008  & 3/2 & 5 & 5888 & 10.18 $\pm$ 0.98 \\
        3.048 & 3/2 & 6  & 6199 & 7.78 $\pm$ 0.31 \\
        3.085 & 1/2 & 7  & 6632 & 4.88 $\pm$ 0.78 \\
        3.091  & 3/2 & 7 & 6558 & 7.28 $\pm$ 1.17 \\
        3.131 & 1/2 & 8  & 7033 & 3.50 $\pm$ 0.51 \\
        3.136 & 3/2 & 8  & 6966 & 6.69 $\pm$ 1.21 \\
        3.179 & 1/2 & 9 & 7483 & 4.18 $\pm$ 0.64 \\
        3.183 & 3/2 & 9 & 7422 & 4.54 $\pm$ 0.83 \\
        3.229 & 1/2 & 10 & 7483 & 2.95 $\pm$ 0.93 \\
        %3.337& 1/2 & 12 & 9126 & 1.25 $\pm$ 0.43 \\
    \end{tabular}
    \caption{(Top) Intensities of the OH lines detected with MIRI-MRS. (Bottom) Intensities of the OH lines detected with NIRSpec. The uncertainties are only the fitting error. The impact of calibration effects is not taken into account and the associated uncertainties can be as high as 20\%.}
    \label{tab:intensity}
\end{table}

\begin{table}
    \centering
    \begin{tabular}{c|c|c|c}
    $\lambda$ ($\mu$m)  & Transition & $E_{\rm up}/k_B$ (K) & $I$ ($\times$10$^{-4}$ erg cm$^{-2}$ s$^{-1}$ sr$^{-1}$) \\
        \hline
    17.035  & 0-0 S(1) & 1015 & 1.55 $\pm$ 0.19 \\
    12.279  & 0-0 S(2) & 1682 & 2.26 $\pm$ 0.30 \\
    9.665 & 0-0 S(3) & 2504 & 12.85 $\pm$ 0.68 \\
    8.025 & 0-0 S(4) & 3474 &	8.70 $\pm$ 2.03	\\
    6.910 & 0-0 S(5) & 4586& 	20.27 $\pm$ 0.12	\\
    6.109 & 0-0 S(6) & 5830 & 	5.30 $\pm$ 0.12 \\
    5.511 &  0-0 S(7) & 7197 &	10.01 $\pm$ 0.20 \\
    5.053 & 0-0 S(8) & 8677	& 1.96 $\pm$ 0.15 \\
    4.695 & 0-0 S(9) &10262 &	3.31 $\pm$ 0.41 \\
    4.410 &  0-0 S(10) & 11940 &	0.52 $\pm$ 0.33 \\
    4.181 & 0-0 S(11) & 13703 &	1.05 $\pm$ 0.16	\\
    3.846 & 0-0 S(13)&17444& 	0.47 $\pm$ 0.07	\\
    3.724 & 0-0 S(14) & 19403 & 0.24 $\pm$ 0.13 \\
    3.626 & 0-0 S(15) & 21412 & 0.15$\pm$ 0.09 \\
    2.627 & 1-0 O(2) & 5987 & 9.89 $\pm$ 3.15	\\
    2.407 & 1-0 Q(1) & 6149 & 35.70 $\pm$ 0.62\\
    2.223 & 1-0 S(0) & 6472 & 7.06 $\pm$ 0.22 \\
    2.122 & 1-0 S(1) & 6952 & 25.03 $\pm$ 0.36	\\
    2.034 & 1-0 S(2) & 7585 & 7.04 $\pm$ 0.20 \\
    1.958 & 1-0 S(3) & 8365 & 14.23 $\pm$ 0.42	\\
    1.892 & 1-0 S(4) & 9287 & 2.87 $\pm$ 0.13  \\
    1.836 & 1-0 S(5) & 10342 & 4.16 $\pm$ 0.19 \\
    1.788 & 1-0 S(6) & 11522 & 0.42 $\pm$  0.17 \\
    1.748 & 1-0 S(7)	& 12818 & 1.06 $\pm$ 0.29	\\
    2.786 & 2-1 O(2) & 11636 & 0.65 $\pm$ 0.13	 \\
    2.974 & 2-1 O(3) & 11790 & 0.64 $\pm$ 0.08\\
    2.356 & 2-1 S(0) & 12095 & 0.20 $\pm$ 0.08 \\
    2.248 & 2-1 S(1)	& 12550 & 0.52 $\pm$ 0.18	\\
    2.073 & 2-1 S(3) & 13891 & 0.47 $\pm$ 0.14 \\
    2.604 & 2-1 Q(5) & 13891 & 0.27 $\pm$ 0.15 	\\
    2.654 & 2-1 Q(7) & 15763 & 0.19 $\pm$ 0.08 \\
    1.853 & 2-1 S(7) & 18107 & 0.28 $\pm$ 0.19 \\
    1.233 & 3-1 S(1) & 17819 & 0.72 $\pm$ 0.10 \\
    2.201 & 3-2 S(3)&19086& 0.24$\pm$ 0.17 \\
    2.066 & 3-2 S(5) & 20857 & 0.16 $\pm$ 0.02\\
    3.376 & 4-3 O(3) & 22080 & 0.10 $\pm$ 0.03 \\
    1.242 & 4-2 S(4) & 24734 & 0.15 $\pm$ 0.09 \\
   \end{tabular}
                         
    \caption{Intensities of the H$_2$ lines detected at the bright spot with the OFF position subtracted. The uncertainties are only the fitting error. The impact of calibration effects is not taken into account and the associated uncertainties can be as high as 20\%.}
    \label{tab:intensity_H2}
\end{table}

\clearpage

\end{document}